\documentclass[11pt]{article}
\usepackage{geometry}[a4paper]
\usepackage{lineno}
\usepackage{authblk}
\usepackage{tikz}
\usepackage{mystuff}
\usepackage{hyperref}

\usepackage{dcolumn}
\usepackage{graphicx} % Required for inserting images
\pgfplotsset{compat=1.18}

\title{\boldmath Multipartite Entanglement Structure of Fibered Link States}

\author[a,b,c,d]{Vijay Balasubramanian}
\author[a]{Charlie Cummings}
\affil[a]{David Rittenhouse Laboratory, University of Pennsylvania,  209 S.33rd Street, Philadelphia, PA 19104, USA}
\affil[b]{Santa Fe Institute, 1399 Hyde Park Road, Santa Fe, NM 87501, USA}
\affil[c]{Theoretische Natuurkunde, Vrije Universiteit Brussel, Pleinlaan 2, B-1050 Brussels, Belgium}
\affil[d]{Rudolf Peierls Centre for Theoretical Physics, University of Oxford,Oxford OX1 3PU, U.K.}

\date{\today}

\begin{document}

\maketitle
%\flushbottom

\paragraph{Abstract}We study the patterns of multipartite entanglement in Chern-Simons theory with compact simple gauge group $G$ and level $k$ for states defined by the path integral on ``link complements'', i.e., compact manifolds whose boundaries consist of $n$ topologically linked tori. We focus on link complements which can be described topologically as fibrations over a Seifert surface. We show that the entanglement structure of such fibered link complement states is controlled by a topological invariant, the monodromy of the fibration.   Thus, the entanglement structure of a Chern-Simons link state is not simply a function of the link, but also of the background manifold in which the link is embedded. 
In particular, we show that any link possesses an embedding into some background that leads to  Greenberger–Horne–Zeilinger state (GHZ)-like entanglement. 
Furthermore, we demonstrate that all fibered links with periodic monodromy have GHZ-like entanglement, i.e., a partial trace on any link component produces a separable state. These results generalize to any three dimensional topological field theory with a dual chiral rational conformal field theory.

\tableofcontents

\section{Introduction}

Quantum entanglement in many-body systems can exhibit rich multipartite structure \cite{walter2017multipartiteentanglement}. For example, a system of just three qubits can display two types of entanglement up to local unitary transformations: the GHZ type, $|\mathrm{GHZ}\rangle = (|000\rangle + |111\rangle) / \sqrt{2}$, where the density matrix is separable after a partial trace on one qubit, and the W type, $|\mathrm{W}\rangle = (|100\rangle + |010\rangle + |001\rangle)/\sqrt{3}$, where the density matrix remains entangled after a partial trace \cite{walter2017multipartiteentanglement}.  Generalized GHZ states like $(|000\cdots 0\rangle + |111 \cdots 0\rangle) / \sqrt{2}$ also have the property that they have the same von Neumann entropy for the density matrix arising from any partial trace. No local unitary transformation can convert a GHZ-like state into a W-like one.\footnote{More generally, these states fall into different equivalence classes under stochastic local operations and classical communication (SLOCC).}   Meanwhile, four qubits can be entangled in nine different ways up to unitary transformation \cite{fourqubits}. In fact,  several of these distinct classes come in continuous families, %that cannot be unitarily transformed into each other, 
so in some sense there are infinitely many patterns of four-party entanglement. Because the dimension of the Hilbert space of $n$ qubits is exponential in $n$, the zoo of entanglement patterns continues to expand rapidly with the number of qubits. 

Multiparty entanglement helps to characterize topological phases of matter \cite{Kitaev_2006}, quantum chaos \cite{Hosur_2016}, topological field theories \cite{Balasubramanian_2017,Balasubramanian_2018,Salton_2017}, and states of strongly coupled many-body systems with geometric dual descriptions \cite{Bao_2015,Balasubramanian_2014,Hubeny_2018,Czech_2023,Akers_2022,balasubramanian2024signalsmultipartyentanglementholography,Melnikov_2023,melnikov2023connectomesholographicstates}.   Multi-party entanglement is also robust to local perturbations of a state, and hence may provide error-correcting information representations for quantum computation, for example through the use of stabilizer codes \cite{Nezami_2020,Bao_2022,Saltonswingle_2017} or states in systems with topological dynamics \cite{freedman2002topologicalquantumcomputation,Kitaev_2003,microsoft2025interferometric,PhysRevLett.133.230603,Melnikov_2019,melnikov2024jonespolynomialsmatrixelements}.

Chern-Simons theories provide one of the simplest settings for studying the patterns  of multiparty entanglement that are naturally produced in the ground states of local Hamiltonians, because the observables are computed in terms of topological invariants of the manifold on which the theory is defined \cite{Witten:1988hf}.  
Specifically, consider a ``link complement'', three-manifolds ${\cal M}$ whose boundaries consist of $n$ topologically linked tori, which we explain in more detail below.  
Famously, the wave function of the link-complement state defined on these boundaries by the Chern-Simons path integral on ${\cal M}$ can be written in terms of topological invariants of the link \cite{Witten:1988hf}. The authors of \cite{Balasubramanian_2017,Balasubramanian_2018} used these methods to show that torus links in $S^3$ (links that can be drawn on the surface of a 2-torus within a 3-sphere) are associated to GHZ-like states, in that a partial trace of any link component produces a seperable state.  These authors also showed in a number of examples that hyperbolic links in $S^3$ (links whose complement admits a complete hyperbolic structure) are associated to W-like states, which remain entangled after a partial trace. 

In this paper, we consider link states of Chern-Simons theory which have the additional property that the link complements are given by circle fibrations over a Seifert surface (see below for definitions).  We show that the entanglement structure of fibered link complement states is controlled by the monodromy of the fibration, and that any link has some embedding into some background manifold that leads to GHZ-like entanglement  (partial traces produce a separable state). Furthermore, we show that all fibered links with periodic monodromy have GHZ-like entanglement.  In the remainder of this introduction we will review Chern-Simons theory and its link states.  Six sections follow.  In Sec.~\ref{sec:math} we discuss the structure of fibered link complements and the monodromy of the fibration.  In Sec.~\ref{sec:linkstates} we explain how to compute Chern-Simon link states.  In Sec.~\ref{sec:trivial} we analyze the entanglement pattern induced in link states in the link complement has trivial monodromy, as well as a formula for link states in the more general case.  In Sec.~\ref{sec:threekinds} we discuss the possible kinds of monodromy, and compute the link states of some examples. Finally, in Sec.~\ref{sec:periodic} we show that periodic mononodromy of a link complement induces GHZ-like entanglement of the associated link state.   We end with a brief conclusion in Sec.~\ref{sec:conclusions}.

\subsection{Chern-Simons theory}
We will now review three-dimensional Chern-Simons theory with gauge group $G$, which we take to be simple and compact. Let $M$ be some closed three manifold without boundary. For concreteness, one could restrict $M$ to be, for example, the three sphere $S^3$, but we will not need to do that. Chern-Simons theory on $M$ is described by the partition function \cite{Witten:1988hf}
\begin{align}
    Z[M] &= \int DA \, e^{i k I_{CS}[A]} 
    ~~~; \nonumber\\
   I_{CS} &= \frac{1}{4\pi} \int d^3 x  \, \Tr[A \wedge dA + \frac{2}{3} A \wedge A \wedge A]  \,.
%   =
%   \frac{1}{4\pi} \int d^3 x \epsilon^{ijk}\Tr[A_i \partial_j A_k + \frac{1}{3} A_i [A_j,A_k]]
%   \nonumber
\end{align}
Here, $k$ is a coupling constant, also called the {\it level} of the theory, $A$ is the gauge field valued in the Lie algebra $\mathfrak{g}$ of $G$, and $\Tr$ is a trace on the Lie algebra,  normalized so that $I_{CS}$ is gauge invariant mod $2\pi$. The Chern-Simons action can be shifted by multiples of $2\pi$ via gauge transformations that are not continuously connected to the identity.  Because of this ambiguity, the level $k$ must be an integer to ensure that $Z[M]$ is well defined. Chern-Simons theory is  \emph{topological}, as the action does not depend on the metric on $M$, even after renormalization. The topological nature of Chern-Simons theory implies that information encoded in states in the Chern-Simons Hilbert space will not be sensitive to local deformations, i.e., to the local physical noise that afflicts quantum computers
 built from non-topological components \cite{freedman2002topologicalquantumcomputation}.

Suppose we cut $M$ along a closed two-dimensional submanifold  $\Sigma \subset M$, a so-called Heegaard splitting of $M$, so that $M = M_1 \sqcup_\Sigma M_2$. We can interpret the Chern-Simons path integral as computing an overlap between states
\begin{align}
    Z[M] = \braket{M_1}{M_2}_\Sigma \,.
\end{align}
The states $\ket{M_1}, \ket{M_2}$ are defined on a Hilbert space $\Ha(\Sigma)$ associated to the splitting surface $\Sigma$. We can understand $\ket{M_1}$ as being prepared via a Hartle-Hawking prescription \cite{hhwavefn} for the path integral $Z[M_1]$ on the manifold-with-boundary $M_1$, with gluing boundary conditions on $\partial M_1 = \Sigma$, and similarly for $\ket{M_2}$. We can think about different states in $\Ha(\Sigma)$ as being prepared by different manifolds $M_i$ such that $\partial M_i = \Sigma$. We are interested in a particular class of such states which we will explain below.

\subsection{Link states}

Let $M$ be a closed three manifold without boundary, and $\La^n$ be an $n$-component link in $M$. Namely, $\La^n$ is an embedding
\begin{equation}
    \La^n: \bigsqcup_{i=1}^n S^1 \into M
\end{equation}
of $n$ disjoint circles into $M$. Such a link defines a three-manifold with boundary called the \emph{link complement}
\begin{equation}
    M(\La^n) = M \setminus \mathcal{N}(\La^n)\,,
\end{equation}
where $\mathcal{N}(\La^n)$ is a tubular neighborhood of the link in $M$. Intuitively, the link complement is constructed by ``drilling out'' $n$ solid tori, linked together according to $\La^n$, from $M$. The boundary of the link complement is
\begin{equation}
    \partial M(\La^n)  = \bigsqcup_{i=1}^n T^2\,,
\end{equation}
i.e., $n$ disjoint copies of the usual 2-torus. The path integral for Chern-Simons theory on this manifold is a functional of the boundary conditions of the gauge field on $\partial M(\La^n)$.
As outlined above, the space of such functionals defines a Hilbert space on $\partial M(\La^n)$, and is known to be the Hilbert space of conformal blocks of the Wess-Zumino-Witten (WZW) field theory of the same gauge group $G$ and level $k$ as the associated Chern-Simons theory \cite{Witten:1988hf,Moore:1988qv}.\footnote{A conformal block is a special function which satisfies the conformal Ward identities. The chiral parts of CFT correlation functions are linear combinations of these special functions. The Hilbert space of these functions (along with a natural inner product between them) is the one dual to Chern-Simons. These details will not be too important for our purposes.} We denote the state prepared by this path integral as $\ket{M(\La^n)}$, and refer to it as a \emph{link state}.

From the perspective of the WZW theory on the disconnected boundary tori, the Hilbert space on $\partial M$  factorizes: 
\begin{equation}
    \Ha(\partial M(\La^n)) = \bigotimes_{i=1}^n \Ha(T^2)\,.
\end{equation}
Because the Hilbert space has a  product structure, we can study the entanglement of such states across arbitrary bipartitions of the tori. In other words, the question we want to ask is how the different boundary tori are entangled to each other.

Let us now discuss the structure of each factor $\Ha(T^2)$. It turns out that each factor has a finite dimensional basis, labeled by the integrable representations of $G$ at level $k$ (see Appendix~\ref{sec:intreps} for more details). This basis is most easily prepared by computing the Chern-Simons path integral on the solid torus $D^2 \times S^1$, with a Wilson loop in the representation $j$ inserted along the non-contractible cycle of the solid torus. This path integral is computed as a functional of the boundary conditions for the gauge field on the boundary $T^2$ of the solid torus. We refer to this state as $\ket{j}$. The set of all such states form orthonormal because of the fusion rules of WZW theory, and span the entire boundary Hilbert space essentially by definition. In the end, what's important is that there is a natural orthonormal basis for the boundary Hilbert space given by
\begin{equation}
    \Ha(\partial M(\La^n)) = \text{span}\left\{ \ket{j_1}\cdots \ket{j_n} | \, j_i \in \text{Reps}(G_k)\right\} \,.
\end{equation}

As a concrete example, if $G=\SU(2)$, then $\text{Reps}(G_k)$ are simply a subset of the usual spins of $\SU(2)$,  $j=0,\frac{1}{2}, \cdots, \frac{k}{2}$. In general, $\text{Reps}(G_k)$ can be thought of as a truncation of the complete list of irreducible representations of $G$; this truncation is what makes the Hilbert spaces finite dimensional. For brevity, we will often abbreviate the natural product basis as 
\begin{equation}
    \ket{J} :=\ket{j_1} \cdots \ket{j_n} \,.
\end{equation}
As we said above, a given link complement $(M,\La^n)$ prepares a specific link state $\ket{M(\La^n)}$ in this Hilbert space. Sometimes, if the background manifold is clear from context, we will write $\ket{\La^n}$ instead.

\section{Link complements and cobordisms} \label{sec:math}

\subsection{Seifert surfaces}

In our analysis, we will need to foliate the link complement with Seifert surfaces.  A Seifert surface $\Sigma_S$ for a link $\La^n$ is a smooth, two-dimensional surface embedded in $M$, with a boundary $\partial \Sigma_S$ isotopic to $\La^n$ in $M$.\footnote{An isotopy is a ``small'' homeomorphism that is continuously connected to the identity. In other words, it is a homeomorphism which does not involve ``ripping'' and regluing, only ambient deformations of $\Sigma_S$.  An non-example is the Dehn twist of the torus.} Such a $\Sigma_S$ exists for any $(M,\La^n)$ \cite{rolfsen1976knots}.  In fact, any link admits many choices of Seifert surface, but up to isotopy, there is a unique minimal genus surface if the link is fibered \cite{kitayama2022surveythurstonnorm} (see below for the definition of fibered). We mean this minimum genus surface whenever we talk about Seifert surfaces below.

As an example, take $\La^n$ be the unknot $0_1$ and $M$ to be $S^3$. In this case, the Seifert surface $\Sigma_S(0_1) = D^2$ is the disk bounded by the unknot, or, more precisely, a disk with boundary on a the surface of a tubular neighborhood of the unknot. A less trivial example is the Seifert surface for the Hopf link $2_1^2$ (two interlinked circles) in the three sphere. To visualize  the Seifert surface is in this case, draw the Hopf link as two concentric circles in a plane with the inner circle twisted around the other one. Then imagine drawing a line segment from one circle to the other and let the endpoints of the segment rotate around both link components at equal speed. The resulting two dimensional surface that is swept out is the annulus with two half-twists \cite{rolfsen1976knots}. $\Sigma_S(2_1^1)$ is therefore homeomorphic to, but not isotopic to, the two holed sphere $\Sigma_{0,2}$, where by $\Sigma_{g,n}$ we mean the Riemann surface of genus $g$ with $n$ boundary components.  The large homeomorphism relating $\Sigma_S$ for the Hopf link to $\Sigma_{0,2}$ is the Dehn twist on the cycle stretching from one boundary to the other which enacts the two required half twists.
%We will return to this Dehn twist momentarily: it will ultimately play an inessential role in what's to come.

\begin{figure}
    \centering
   \includegraphics{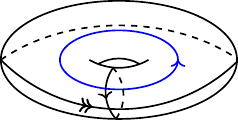}
    \caption{Labeling each cycle of the tubular neighborhood. The blue cycle is the original link component that the tubular neighborhood surrounds. We call the single arrow the $b$ cycle, and the double arrow the $a$ cycle. The cycle with the single arrow parameterizes the $\theta$ coordinate of the link complement. }
    \label{fig:abcycle}
\end{figure}

%The fact that $\Sigma_S$ always exists for any $M(\La^n)$ has a dramatic implication for the topology of link complements. To see why, 
Suppose we embed a Seifert surface $\Sigma_S$ in the link complement $M(\La^n)$. %, such that $\partial \Sigma_S$ has support on $\partial M$. 
This surface will intersect the tori of $\partial M$ along some longitude, which we can call the $a$-cycle of each torus. We can give $M$ coordinates where $\Sigma_S$ is embedded along $(x_\Sigma,y_\Sigma,\theta=0)$. We imagine that $x_\Sigma,y_\Sigma$ run ``along'' the Seifert surface, while $\theta$ runs ``around'' the $b$-cycle of every torus of the link complement. Note that $\theta$ is \emph{not} the coordinate of the link component; see Fig.~\ref{fig:abcycle}.
In other words, if we think of the link component itself as running along the $a$-cycle of each torus, $\theta$ is an angle along the $b$-cycle. 
We can generically extend these coordinates to an open neighborhood of $\Sigma_S$ in $M(\La^n)$. We do so by extending a vector field $\partial_\theta$ into $M$, transverse to $\Sigma_S$, and using this vector field to generate a flow which defines the neighborhood. The picture to keep in mind is that $\Sigma_S$ is ``sweeping out'' a subset of $M(\La^n)$, and the value of $\theta$ is a time parameter for the flow. This procedure defines a foliation of $M(\La^n)$ with leaves isotopic to $\Sigma_S$.  If such a flow can be chosen so that the open neighborhood covers the entire link complement $M(\La^n)$, we say $\La^n$ fibers in $M$, and call $\La^n$ a fibered link.

To visualize this flow, let $\La^n$ be a fibered link. Then, slicing open the link complement on a Seifert surface yields a product manifold:
\begin{equation}
    M(\La^n) \setminus \Sigma_S\cong\Sigma_S \times I\,,
\end{equation}
where $I$ is the interval. To recover $M(\La^n)$, we glue the two endpoints of the interval back together using a map $f: \Sigma_S \to \Sigma_S$. In other words, we identify
\begin{equation}
    \Sigma_S \times \{1\} \mapsto f(\Sigma_S) \times \{0\}\,. \label{eqn:monodromyidentification}
\end{equation}
By definition, $M(\La^n)$ has this form if $\La^n$ is fibered. Thus, for fibered links, $M(\La^n)$ has the form of a trivial cobordism between Seifert surfaces, up to an identification via $f$ (see Fig.~\ref{fig:fibration}). More general links can also be thought of as cobordisms between Seifert surfaces, but in general the cobordisms will have a more intricate bulk topology \cite{morsenovikov, severino2024dislocationsfibrationstopologicalstructure}. Indeed, we can always Heegaard split $M(\La^n) \to M(\La^n) \setminus \Sigma_S$ whether or not $\La^n$ is fibered, and this Heegaard splitting will define a cobordism from $\Sigma_S \to \Sigma_S$. A link is fibered if and only if this cobordism is trivial (does not involve topology changes along the flow) up to the identification map $f$.

\subsection{Non-fibered links} \label{sec:nonfibered}

Fibered links form a wide class of links. In fact, at least when $M=S^3$,  even if a link $\La^n$ is \emph{not} fibered,  there is always a $n+1$ component link $\La^n \cup K$ which \emph{is} fibered in $S^3$, and contains $\La^n$ as a sublink \cite{stallings}.\footnote{See Appendix \ref{sec:whenfibered} for an explanation. It is possible that this also holds for more general background manifolds $M$, but we could not find a reference.}
Furthermore, the extra component $K$ can always be taken to be an unknot. Thus, the method we are about to develop can also be used for non-fibered link states through the following algorithm:
\begin{enumerate}
    \item Construct the larger link $\La^n \cup K$.
    \item Compute the link state of $(M,\La^n \cup K)$ using its fibration data (which we explain how to do below).
    \item Project out the extra component to the trivial representation. This removes the extra component, returning the link state of just $\La^n$, the original link. Up to normalization, the link state for the non-fibered link will be
\end{enumerate}
\begin{equation}
    \ket{M(\La^n)} = [\Id_{\La^n} \otimes \bra{0}_K ]\ket{M(\La^n \cup K)}\,.
\end{equation}
Strictly speaking,  results of this paper about the entanglement structure of non-fibered link states will only hold for the larger link $(M,\La^n \cup K)$. Projecting out the additional component with $\Id_{\La^n} \otimes \bra{0}_K$ will generally change the entanglement structure, though such an operation can only decrease the entanglement between the boundary tori. 
%On the knot theory side, there is no guarantee e.g. that $M(\La^n \cup U)$ is hyperbolic, even if $M(\La^n)$ is. <== turns out this is false, by quantum fibered link monodromy paper
We leave investigation of the effects of this projection on the entanglement structure and the relation to link topology for future work.

\subsection{Fibered link complements}

For the remainder of of this paper, we restrict to fibered links, so $M(\La^n) \setminus \Sigma_S$ has the topology $\Sigma_S \times I$. We review tools for recognizing when a link is fibered in Appendix \ref{sec:whenfibered}. In fact all fibered links of sufficiently low crossing number have been classified \cite{Gabai1986}, and they form a broad class.

%Every torus link in $S^3$ is fibered, but there are examples of fibered hyperbolic links as well. 

To recover the link complement $M(\La^n)$ from the cobordism picture, we explained that we must identify the two ends of the cobordism via a homeomorphism $f: \Sigma_S \to \Sigma_S$ from the surface to itself. Such a map $f$ is called the \emph{monodromy} of the link, and  measures how the Seifert surface is shifted as it twists around the link.  Because $\Sigma_S$ has genus $g$ and $n$ boundary components, it is homeomorphic to $\Sigma_{g,n}$, the genus $g$ Riemann surface with $n$ contractible boundary components. But it is \emph{not} necessarily isotopic to $\Sigma_{g,n}$: the Hopf link is a counterexample, as explained above. That being said, there always exists a homeomorphism $\phi$ such that $\Sigma_S = \phi(\Sigma_{g,n})$. We will find it convenient to think of $\Sigma_S$ this way, as $\Sigma_{g,n}$ provides a canonical ``reference surface'', from which we can obtain $\Sigma_S$ by applying $\phi$. $\phi$ will turn out to play an inessential role in the entanglement entropy of $\ket{M(\La^n)}$.

Clearly, any other map $f'$ related to $f$ by an isotopy of the surface leads to a homeomorphic manifold $M$.  We should really only keep track of $f$ up to isotopy, because the Chern-Simons theory is a topological field theory and its states are invariant under local deformations of link complement.\footnote{This is true up to a choice of framing, as we pointed out above, see \cite{Witten:1988hf}.} Furthermore, we should demand that $f|_{\partial \Sigma_S} = \Id$, because the Seifert surface does not change close enough to the link itself \cite{mappingclass}.
We will denote this equivalence class of maps by $[f]$. The monodromy defined by $[f]$ is a link invariant. The set of such equivalence classes, together with the composition of such maps, defines the \emph{mapping class group} $\text{Mod}(g,n)$, where $g$ is the genus of $\Sigma_S$ and $n$ is the number of link components. We can think of $\text{Mod}(g,n)$ as the set of ``large'' diffeomorphisms of $\Sigma_S$. This is isomorphic to the set of large diffeomorphisms of $\Sigma_{g,n}$, so we will not distinguish them.
%Changing $[f]$ still defines a three-manifold with boundary $(T^2)^n$, but then $M \sqcup (D^2 \times S^1)^n$ (i.e., filling the knot back in with empty space) will generically no longer be homeomorphic to $S^3$. 

\begin{figure}
    \centering
    \includegraphics[]{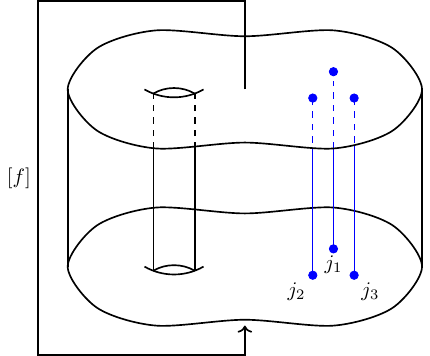}
    \caption{The link complement of a fibered link. In this example, we took $\Sigma_S \cong \Sigma_{1,3}$, and we have glued in the $J$ punctures along the $b-$cycles of each boundary torus. }
    \label{fig:fibration}
\end{figure}

\begin{figure}
    \centering
    
   \includegraphics[width=0.75\textwidth]{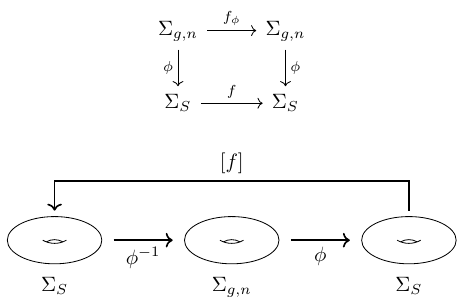}
    \caption{Top: Commutative diagram explaining why the topology of the link complement is independent of $\phi$. The fact this diagram commutes implies that $f$ with base $\Sigma_S$ and $f_\phi$ with base $\Sigma_{g,n}$ produce homeomorphic link complements.
    Bottom: The geometric picture associated with this commutative diagram. The bottom path of the commutative diagram, associated with Eq.~\ref{eqn:fibration_withphi}, corresponds to defining the twisted product with respect to the first or third Riemann surfaces. The monodromy in these cases is $f \circ \phi\circ \phi^{-1}  $ or $\phi \circ \phi^{-1} \circ f$, respectively. Both are equivalent to $f$. The top path of the commutative diagram, associated with Eq.~\ref{eqn:gntwisted}, corresponds to defining the twisted product with respect to the second Riemann surface. The monodromy in this case is $f_\phi$. Because the overall fibration is the same in all three cases, Eq.~\ref{eqn:fibration_withphi} and Eq.~\ref{eqn:gntwisted} represent precisely the same link complements. }
    \label{fig:cd}
\end{figure}

Putting all this together, any fibered link complement has the form of a twisted product 
\begin{align}
    M(\La^n) = \phi(\Sigma_{g,n}) \times_f S^1 \,. \label{eqn:fibration_withphi}
\end{align}
This decomposition of $M(\La^n)$ is almost what we want, but there is a further convenient simplification. Because the diagram of Fig.~\ref{fig:cd} commutes, we can define a new monodromy via $\phi$-conjugation 
\begin{equation}
    [f_\phi] := [\phi^{-1} \circ f \circ \phi]\,,
\end{equation}
and instead take the decomposition
\begin{align}
    M(\La^n) = \Sigma_{g,n} \times_{f_\phi} S^1 \,. \label{eqn:gntwisted}
\end{align}
Whereas $f$ is a map from $\Sigma_S \to \Sigma_S$, $f_\phi$ is a map from $\Sigma_{g,n} \to \Sigma_{g,n}$. Even though $\phi$ is not itself a mapping class group element because it does not fix $\partial \Sigma_{g,n}$ pointwise, $f_\phi$ is still a mapping class group element conjugate to $f$.  To see why, consider the commuting diagram in Fig.~\ref{fig:cd}.  In this diagram $f_\phi$ maps between two copies of $\Sigma_{g,n}$ obtained by applying $\phi^{-1}$ to copies of the Seifert surface $\Sigma_S$. The applications of $\phi^{-1}$ moves boundary points of $\Sigma_S$ to boundary points of $\Sigma_{g,n}$ in the same way.  So $f_{\phi}$ will continue to act as the identity on these points since $f$ did so also. A more precise derivation of this fact uses the fact that $f \mapsto f_\phi$ is an automorphism of the mapping class group. It is shown in \cite{Ivanov1988} that as long as $n \geq 1$, the only non-trivial outer automorphism of the mapping class group is orientation reversal. As $\phi$ is orientation preserving, this means that $f \mapsto f_\phi$ must be an inner automorphism, and so $f_\phi = \widetilde{\phi}^{-1} \circ f \circ \widetilde{\phi}$ for some genuine mapping class group element $\widetilde{\phi}$. 
However, two fibered link complements with conjugate monodromy like $[f]$ and $[f_\phi]$ are homeomorphic as manifolds with boundary \cite{mappingclass}. Thus, because the Chern-Simons path integral is invariant under background homeomorphisms, we can freely replace $[f_\phi]$ with $[f]$, since they are conjugate, and instead consider the homeomorphic decomposition
\begin{align}
    M(\La^n) \cong \Sigma_{g,n} \times_{f} S^1 \,.
\end{align}
This decomposition of $M(\La^n)$ is crucial in what follows. 
The transformation from $\phi(\Sigma_{g,n})$ to $\Sigma_{g,n}$ is quite radical from a three dimensional point of view.
Specifically, we are saying that
%note that the independence of $\phi$ means 
the different link complements
\begin{align}
    M_1(\La_1^n) &= \phi(\Sigma_{g,n}) \times_f S^1 \label{eqn:phi_ind1}\\
    M_2(\mathcal{L}_2^n) \label{eqn:phi_ind2} &=  \Sigma_{g,n} \times_{f} S^1
\end{align}
are homeomorphic, and thus lead to the same Chern-Simons path integral. Although we just showed that $M_1(\La_1^n) \cong M_2(\La_2^n)$, for the moment we will distinguish them. $M_1(\La_1^n)$ is the link complement that we have been calling $M(\La^n)$, while $M_2(\La_2^n)$ is the simplified link complement we will use for computations. 
%For instance, we will see an example in which the components of $\mathcal{L}^n_2$ turns out be unlinked, but embedded in a background with non-trivial topology.  
We will now determine the precise relationship between the link states $\ket{M_1(\La_1^n)}$ and $\ket{M_2(\La_2^n)}$
of these two homeomorphic link complements.

The background manifold $M_1$ is defined by gluing solid tori along the cycles of $\partial M_1(\La_1^n)$ that run along $\partial \phi(\Sigma_{g,n})$, while $M_2$ is defined by gluing the solid tori along $\partial \Sigma_{g,n}$ instead. $\La^n_1$ and $\La^n_2$ are then defined by the linking configuration of these tori  within $M_1,M_2$, respectively. With these definitions, $M_1 \neq M_2$ and $\La^n_1 \neq \mathcal{L}^n_2$ in general. Despite this,  $M_1(\La^n_1)$ and $M_2(\La^n_2)$ are homeomorphic as \emph{link complements} no matter what $\phi$ is. The surprising feature is that each $a$-cycle of $M_2(\mathcal{L}_2^n)$ is just the boundary of $\Sigma_{g,n}$, and from the perspective of $M_2$, the components of $\partial \Sigma_{g,n} = \La_2^n$ are unlinked from each other!\footnote{They are unlinked because we chose our reference surface $\Sigma_{g,n}$ to have contractible boundaries, and $[f]$ acts as the identity at the boundary of the Seifert surface. They can, however, link with topological features of $M_2$ itself.}

Because $M_1(\La_1^n)$ and $M_2(\La_2^n)$ are homemorphic, the path integral on these manifolds is the same.  That means that if we define the same boundary conditions on $\La_1^n$ and $\La_2^n$, the path integral will give the same number.  A natural boundary condition is to set a Wilson line along the $a$-cycle of the boundary tori because this is what defines the link in the first place.  But the $a$-cycles of $\La_1^n$ and $\La_2^n$ will generally be related by unitary transformations local to each boundary component. Geometrically, as long as we fix the gluing which defines $M_1$ vs $M_2$ (e.g., identity, $S$ twist, etc), we could fill in the link components with solid tori in any order we please and obtain the same link state. At the quantum level, the choices of gluing are unitaries, and they are local because they can be applied independently and in any order at each torus boundary. This means in turn that the  the link states $\ket{M_1(\La_1^n)}$ and $\ket{M_2(\La_2^n)}$ will differ in general by a unitary transformation that is local to each boundary component.
%Because $M_1(\La_1^n)$ and $M_2(\La_2^n)$ only differ by our conventions for the gluing of solid tori along their boundaries, the link states $\ket{M_1(\La_1^n)}$ and $\ket{M_2(\La_2^n)}$ can at most differ by a In a unitary transformation that is local to each boundary component. 

Let $\ket{J}_1$ be the (product) basis of $\Ha(T^2)^{\otimes n}$ which corresponds to gluing the Wilson lines parallel to the $a$-cycles of $M_1(\La_1^n)$, and similarly define an analogous basis $\ket{J}_2$ for the $a$-cycles of $M_2(\La_2^n)$.  This local unitary only depends on $\phi$ because $M_1(\La_1^n)$ is produced from $M_2(\La_2^n)$ by application of $\phi$. Thus, we can relate the natural basis for each link complement via 
\begin{equation}
    \ket{J}_1 = \mathcal{U}_\phi  \ket{J}_2 \,, \label{eqn:phirelation}
\end{equation}
where $\mathcal{U}_\phi$ is a local unitary. We can use the same unitary to relate the states produced by the path integral on the link complements, because these are defined on the same Hilbert space:
\begin{equation}
    \ket{M_1(\La_1^n)} = \mathcal{U}_\phi \ket{M_2(\La_2^n)} \,.\label{eqn:m1l1m2l2}
\end{equation}
We argued above that the unitary $\mathcal{U}_\phi$ enacting the change of basis between the boundary tori of the two link complements must be local.
Thus, because $\mathcal{U}_\phi$ is a local unitary, $\ket{M_1(\La_1^n)}$ and $\ket{M_2(\La_2^n)}$ will have precisely the structure of entanglement between the link components. But $\ket{M_2(\La_2^n)}$ is totally independent of $\phi$, so we have just demonstrated that $\phi$ does not affect the entanglement entropies of its associated link state. This is a generalization of the observation of \cite{Balasubramanian_2017} that the entanglement entropy of link states is independent of the framing of $\La^n$ within $M$ for more complicated Dehn fillings of $M(\La^n)$, i.e. different ways of filling tori into the link complement to make the background manifold. More generally, every R\'enyi entropy of a given link state is independent of $\phi$.  We will analyze the structure of $\mathcal{U}_\phi$ in more detail in Sec.~\ref{sec:Uphireturn}.

As a concrete example, consider the Hopf link $2^2_1$ in $S^3$. Removing one link component, we see that $S^3 \setminus \mathcal{N}(0_1) \cong D^2 \times S^1$, where $\mathcal{N}(0_1)$ is the tubular of one link component. This is just the usual picture of $S^3$ as a gluing of two solid tori. To remove the second link component, we drill out another solid torus within $D^2 \times S^1$ along the $S^1$, giving $\Sigma_{0,2} \times S^1$. But this is the same link complement we would get from removing two unlinked (from each other) circles from the non-contractible cycle of $S^2 \times S^1$. Thus, the pairs $$(S^3,2^2_1) \cong (S^2 \times S^1, 0_1 \sqcup 0_1)$$ produce homeomorphic link complements. In Sec.~\ref{sec:hopflink}, we will confirm explicitly that these link complements produce link states that only differ by a local unitary. 

The reason this is not immediately intuitive is that the homeomorphism relating $M_1(\La_1^n)$ and $M_2(\La^n_2)$ is not an isotopy, and so they can look quite different when embedded in $\R^3$. The fact that the topology of the link complements independent of $\phi$ means there are an {\it infinite} number of pairs $(M,\La^n)$ which produce the same link state, up to local unitaries.  On the one hand, it is expected that the entanglement entropy of a link state does not depend on $\phi$, because it should only depend on $M(\La^n)$ up to homeomorphism. On the other hand, this is surprising, because $\phi$ is essential in making sure that the boundary of the Seifert surface is the link $\La^n$ in question.

What we are finding is that the R\'enyi spectrum of a given link state is actually determined by the other topological data $(g,n,[f])$. Because of this, it is convenient to use the notation
\begin{align}
    \ket{M_1(\La_1^n)} &\mapsto \ket{M(\La^n)}\,, \\
    \ket{M_2(\La_2^n)} & \mapsto \ket{g,n,[f]}\,.
\end{align}
In this notation, \eqref{eqn:m1l1m2l2} reads
\begin{equation}
    \ket{M(\La^n)} = \mathcal{U}_\phi \ket{g,n,[f]}\,.
\end{equation}
%Because we are interested in the multipartite entanglement patterns of link states in Chern-Simons theories, we will mostly focus on $\ket{g,n,[f]}$ rather than $\ket{M(\La^n)}$ itself in this paper. The above discussion shows that there is no loss of generality in doing so. 
Finally, we note in passing that there is not necessarily an easily computable \footnote{If this map \emph{were} simple, then this would contradict the results of \cite{agol2004computationalcomplexityknotgenus}, which show the calculation of a knot's genus is NP-complete.} map 
\begin{equation}
    (M,\La^n) \mapsto (g,n,[f]) \label{eqn:phitimesgnf}
\end{equation}
from a choice of background/link to fibration data. The tuple $(g,n,[f])$ is called an \emph{open book decomposition} of $(M,\La^n)$. Nevertheless, the algorithm described above provides such a map in principle, and is well defined up to isotopy because each element of $(g,n,[f])$ is a topological invariant. 
This map is surjective, as we can always glue solid tori back into the fibration defined by $ (g,n,[f])$ via the identity (or any other gluing $\phi$ we like). This produces \emph{some} background $M$, and removing the same tori we glued in will then define \emph{some} link $\La^n$ in that background manifold.

\section{Computing the link states}
\label{sec:linkstates}

The link state $\ket{M(\La^n)}$ is prepared via the Chern-Simons path integral on $M(\La^n)$. If the link $\La^n$ fibers in $M$, then as we explained above, we can decompose the link complement as
\begin{equation}
    M(\La^n) \cong \Sigma_{g,n} \times_{f} S^1\,, 
    \label{eqn:fibration}
\end{equation}
where $\Sigma_{g,n}$ is the genus $g$ surface with $n$ boundary components, and $[f]$ is an element of the mapping class group $\text{Mod}(g,n)$.

Previous work \cite{Balasubramanian_2017,Balasubramanian_2018} noted that $\ket{M(\La^n)}$ can be expanded in the product basis $\ket{J} = \ket{j_1} \cdots \ket{j_n}$. This is the basis of cycles that are parallel to $\partial \Sigma_S = \partial (\phi(\Sigma_{g,n}))$. With this basis, the unnormalized\footnote{For convenience, and to reduce clutter, we will often work at the level of unnormalized states. There is no ambiguity, because we could always just divide by a normalization constant.} link state is given by 
\begin{equation}
    \ket{M(\La^n)} = \sum_J V_J(\La^n)^* \ket{J}\,, \label{eqn:acycle}
\end{equation}
where $V_J$ is the $J$-colored Jones polynomial of the link $\La^n$ in $M$. However, in principle, we can expand $\ket{M(\La^n)}$ in any basis we like, and a different choice will allow us to exploit the fibration structure more explicitly. 
%We will expand $\ket{M(\La^n)}$ in the basis of cycles which run along the $S^1$ of the fibration of \eqref{eqn:fibration}. 

\subsection{Picking the right basis}

To construct the proper basis for expanding $\ket{M(\La^n)}$, we first consider the state $\ket{g,n,[f]}$ defined in the previous section. For each boundary torus, let the basis $\ket{J}_2$ be the one which runs parallel to $\partial \Sigma_{g,n}$, as above. Thinking of $\ket{J}_2$ as the $a$-cycles, the  conjugate $b$-cycles have linking number one with the $a$-cycles. These $b$-cycles run ``orthogonally" to $\Sigma_{g,n}$ in the link complement, and so run along the $S^1$ in the fibration of \eqref{eqn:fibration}.   On each boundary torus, the $a$-cycle and $b$-cycle basis are related by the modular transformation $S$ which sends the modulus of the torus $\tau \to -\frac{1}{\tau}$.  We are going to find that the $b$-cycle basis is more convenient for computations. To this end, take $S_1$ and $S_2$ to be the S-transformations acting on the $\ket{J}_1$ and $\ket{J}_2$ bases, which are the $a$-cycle bases for $\partial \Sigma_S$ and $\partial \Sigma_{g,n}$, respectively. %So $S_2$ is defined such that
%\begin{equation}
%    \bra{j}_1 S_1 \ket{\ell}_1 = \bra{j}_2 S_2 \ket{\ell}_2 \,.
%\end{equation}
%where $S$ is the natural choice of modular transformation in the cycles in the $\ket{J}$ basis. \VB{Can we simplify or remove the last couple of sentences.}
By \eqref{eqn:phirelation},
\begin{equation}
    S_1^{\otimes n} = \mathcal{U}_\phi \, S_2^{\otimes n} \, \mathcal{U}_\phi^\dagger \,. \label{eqn:sameS}
\end{equation}

We will now expand $\ket{g,n,[f]}$ in the $b$-cycle basis $(S_2)^{\otimes n} \ket{J}_2$. Geometrically we can think of this as gluing the Wilson loops $\ket{J}_2$ along the $b$-cycles of the boundary of the link complement. 
We can see from the gluing procedure above (also see Fig.~\ref{fig:fibration}) that the overlap of $\ket{g,n,[f]}$ with this basis is by definition
\begin{equation}
    \bra{g,n,[f]} S_2^{\otimes n}  \ket{J}_2 = Z\left[\Sigma_{g,J} \times_f S^1\right]\,,
    \label{eq:overlap1}
\end{equation}
where $Z$ is the Chern-Simons path integral  and $\Sigma_{g,J}$ refers to the genus $g$ surface with defects labeled by representations $J = j_1, \cdots, j_n$.
%, which give boundary conditions for the path integral. 
This is because, after gluing in the solid tori to compute the overlap on the left side of \eqref{eq:overlap1}, the $S$ transformations place the Wilson lines along the ``time'' direction of the $S^1$.   As a result they appear as point defects on the genus $g$ Riemann surface.\footnote{In this paper, we will not distinguish between punctures, marked points, and defects.}

Using \eqref{eqn:phirelation} and \eqref{eqn:phitimesgnf}, we can relate this to the amplitude of $\ket{M(\La^n)}$ given by
\begin{equation}
    \bra{M(\La^n)} \mathcal{U}_\phi \,    S_2^{\otimes n} \, \mathcal{U}_\phi^\dagger \ket{J}_1 = Z\left[\Sigma_{g,J} \times_f S^1\right]\,.
\end{equation}
However, notice that by \eqref{eqn:sameS}, this is the same thing as
\begin{equation}
    \bra{M(\La^n)} S_1^{\otimes n} \ket{J}_1 = Z\left[\Sigma_{g,J} \times_f S^1\right]\,. \label{eqn:partfnbcycle1}
\end{equation}
For convenience, we will now define $\mathcal{S} = S_1^{\otimes m}$, leaving the specific power $m$ to be implicit from the Hilbert space $\mathcal{S}$ is acting on. In this case, $m=n$. Furthermore, we will drop all the subscripts on $\ket{J}_1$, in favor of a more streamlined notation $\ket{J}$. The simple form of this overlap shows that it will be useful to expand $\ket{M(\La^n)}$ in the $\mathcal{S}\ket{J}$ basis as
\begin{align}
    \ket{M(\La^n)} =  \sum_J Z\left[\Sigma_{g,J} \times_f S^1\right]^* \mathcal{S}\ket{J}\,. \label{eqn:partfnbcycle2}
\end{align}
We will now compute this partition function explicitly using canonical quantization.

\subsection{Canonical quantization}

The fibration structure over the circle $S^1$ can be exploited to compute the partition function via canonical quantization \cite{Witten:1988hf}. As usual in quantum field theory, the overall circle implies we can compute the partition function as the trace of the operator which enacts ``time evolution'' along it. The global ``time evolution'' in this case is the monodromy $[f]$, and it acts on the genus $g$ defect Hilbert space $\Ha(\Sigma_{g,J})$ \cite{Witten:1988hf}. This is \emph{not} the boundary Hilbert space $\Ha(T^2)^{\otimes n}$, but the Hilbert space of the ``spatial'' slice $\Sigma_{g,J}$ of the fibration, which we elaborate upon below. Now let $K_f$ be the unitary operator which represents the monodromy $[f]$ on this Hilbert space. The desired partition function can then be computed as 
\begin{equation}
    Z\left[\Sigma_{g,J} \times_f S^1\right] = \tr(K_f^\dagger)_{\Ha(\Sigma_{g,J})} \,. \label{eqn:partfnbcycle3}
\end{equation}
The path integral computes $\tr(K_f^\dagger)$, as opposed to $\tr(K_f)$, because of our convention for the monodromy in Eq.~\ref{eqn:monodromyidentification}. $f$ is defined by gluing $\Sigma_S \times \{1\}$ to $\Sigma_S \times \{0\}$, which is equivalent to gluing $\Sigma_S \times \{0\}$ to $\Sigma_S \times \{1\}$ using $f^{-1}$. Thus, the partition function computes the trace of the operator which sends states on $\Sigma_S$ to $f^{-1}(\Sigma_S)$, or $K_{f^{-1}}$. Because $K_f$ is a unitary representation of the mapping class group, this is the same as $K_f^\dagger$. While we could eliminate this dagger by changing our convention, we opt to maintain it for two reasons. First, our convention is more common in the math literature, and so makes referencing the monodromy of known fibered links easier. Second, this dagger eliminates extra complex conjugates in some later equations, making them slightly cleaner.
For brevity, we will now write $\Ha(g,J)$ for $\Ha(\Sigma_{g,J})$, and $\tr(K_f[J])$ for $\tr(K_f)_{\Ha(\Sigma_{g,J})}$. Returning to the problem at hand, we can combine \eqref{eqn:partfnbcycle2} and \eqref{eqn:partfnbcycle3} to see that
\begin{align}
    \ket{M(\La^n)} = \mathcal{S} \sum_J \tr(K_f[J]) \ket{J}\,. \label{eqn:bcycle}
\end{align}
where we used that $\tr(K_f[J]^\dagger)^* = \tr(K_f[J])$. 
%The basis $\ket{J}$ implicitly depends on $\phi$ via \eqref{eqn:phirelation}, so varying $\phi$ will simply change $\mathcal{S}_\phi$.

Note that we can compare the coefficients of $\ket{M(\La^n)}$ in the basis expansions \eqref{eqn:acycle} and \eqref{eqn:bcycle} to see that
\begin{equation}
    V_J(\La^n)^* = \sum_L  \mathcal{S}_{JL} \tr(K_f[L]) \label{eqn:jonesmonodromy}
\end{equation}
This (anti)linear relationship between the colored Jones polynomials $V_J(\La^n)$ and link monodromy $\tr(K_f[J])$ holds for any fibered link. So one interpretation of our methods is as an efficient way to compute colored Jones polynomials for a link $\La^n$ based on the topological invariants $(g,n,[f])$.

\subsection{The genus $g$, $J$-defect Hilbert Space} \label{sec:pants}

As is beautifully explained in \cite{Witten:1988hf}, the Hilbert space $\Ha(g,J)$ has a convenient basis which can be identified the conformal blocks of the associated WZW model with gauge group $G$ and level $k$. The structure of this Hilbert space is extensively discussed in \cite{Moore:1988qv}, as WZW models are a special case of rational conformal field theories (RCFTs). In particular, this Hilbert space is finite dimensional for every surface $\Sigma_{g,J}$. One can build this $\Sigma_{g,J}$ Hilbert space out of smaller pieces via sewing procedures outlined in \cite{Moore:1988qv}. Here, we briefly explain how this sewing  works.

\begin{figure*}
    \centering
    \includegraphics[]{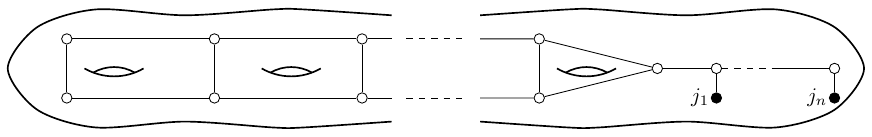}
    \caption{A pair of pants decomposition for $\Sigma_{g,J}$. The defects $J$ are the representations we glue to the boundary components of $\Sigma_{g,n}$ to create $\Sigma_{g,J}$.
    Each dot represents thrice punctured sphere, and each leg represents a gluing of the punctures. A vertex with three legs is a pair of pants: a vertex with fewer than three legs can be viewed as a pair of pants that has been ``capped off'' with the trivial representation (i.e., the puncture has been removed). Each black dot is a puncture that have been left behind. One can think of this as gluing in a disk with a marked point of the relevant representation. }
    \label{fig:pants}
\end{figure*}

Decompose $\Sigma_{g,J}$ into some pair of pants decomposition. Any will do, but we can, for example, take the decomposition outlined in Fig.~\ref{fig:pants}. Each pair of pants is homeomorphic to a 3-punctured sphere. Each 3-punctured sphere (with reps $ijk$ at the punctures) has a Hilbert space $V^j_{ik}$ associated to it, where $i,j,k$ are representations of $G$ at level $k$. We think of $ik$ as flowing ``into'' the sphere and $j$ as flowing ``out of'' the sphere. The entire Hilbert space $\Ha(g,J)$, then, is the tensor product over the pairs of pants, direct summed over all the possible representations at the sewed punctures, with the constraint that the representations at each point agree when they are glued. The remaining punctures labeled $J$ are held fixed, and those that are ``capped off''  (see Fig.~\ref{fig:pants}) are given the trivial representation.

This algorithm produces a Hilbert space for every pair of pants decomposition of $\Sigma_{g,J}$.
It is non-trivial that the resulting Hilbert space is independent of the particular decomposition.
%so a priori there is no reason that all possible Hilbert spaces obtained in this way should agree. 
The fact that all decompositions span  the same Hilbert space is a consequence of \emph{duality} as defined in \cite{Moore:1988qv}. Of course, the explicit bases produced by different decompositions may not be same. The transformations relating each basis are called duality transformations.

\section{Trivial and nontrivial monodromy} 
\label{sec:trivial}

%Now that we understand the overall structure of the Hilbert space, we can use this to compute its dimension. 

\subsection{The case of trivial monodromy}

First, let us us consider fibered links whose complements have trivial monodromy, $[f] = \Id$. 
As we showed in \eqref{eqn:bcycle}, the link state is controlled by the trace of the monodromy, which in this case is  $\tr(\Id)_{\Ha(g,J)} = \dim(\Ha(g,J))$. 
%\begin{align}
%    \tr(\Id)_{\Ha(g,J)} = \dim(\Ha(g,J)) \,.
%\end{align}
So we are calculating the (unnormalized) link state
\begin{align}
    \ket{g,n,[f]=\Id} = \mathcal{S}\sum_J \dim(\Ha(g,J)) \ket{J}\,. \label{eqn:triviallinkstate}
\end{align}
%Because $\dim(\Ha(g,J)) = \tr(\Id)_{\Ha(g,J)}$, this computation will be a warm-up for the more general case.
When computing $\dim(\Ha(g,J))$, it is convenient to define the fusion matrices $N^j_{ik} = \dim(V^j_{ik})$, where $V^j_{ik}$ is the Hilbert space associated to the three-punctured sphere with representations $i,j,k$ at the punctures. The fusion coefficients have a well known form due to Verlinde \cite{Verlinde:1988sn}:
\begin{align}
    N^j_{ik} = \sum_\ell \frac{S_{k\ell} S_{i\ell} (S^\dagger)_{j\ell}}{S_{0\ell}} \,. \label{eqn:verlinde}
\end{align}
Here, $S_{ij}$ is the matrix element of the modular transformation $S$ which also appeared above. Some crucial properties of these matrix elements are \cite{DiFrancesco:1997nk,Shi_2020}:
\begin{align}
    S_{ab} &= S_{ba} \\
    S_{ab}^* &= S_{\bar{a} b} \\
    \sum_c S_{ac}S_{cb} &= \Theta_{ab} \label{eqn:Ssym}
\end{align}
where $\bar{a}$ is the dual representation (charge conjugate) to $a$, and $\Theta$ is the charge conjugation operator. In particular, we note that because $\bar{0} = 0$, $S_{0\ell}$ is real for any $\ell$.

As described above, a basis for the Hilbert space $\Ha(g,J)$ can be associated with a pair of pants decomposition of $\Sigma_{g,J}$. For every pair of pants, we get a factor of $V^j_{ik}$: when computing the trace of the identity, this gives a factor of $\dim(V^j_{ik}) = N^j_{ik}$. The tensor product over each $V^j_{ik}$ means we multiply each factor of $N^j_{ik}$ together, and the direct sum over representations at each puncture means we sum over all index contractions. For $\Sigma_{g,J}$, we use the pair-of-pants decomposition given in Fig.~\ref{fig:pants} to compute this. In this decomposition, we use $2g+1$ pairs-of-pants to build the ``genus-$g$ part'' of the surface, and $n$ pairs-of-pants to construct the ``punctures-$J$ part''. Each pair of pants contributes a factor of $(S_{0\ell})^{-1}$ from the denominator of the Verlinde formula. Because $S$ is unitary, all contractions of $S,S^\dagger$ cancel. Additionally, in this decomposition, two extra punctures must be capped off of the genus-$g$ part to get the desired surface, and one from the punctures-$J$ part: this contributes an additional factor of $(S_{0\ell})^{3}$. Finally, each free index $j$ from the punctures $J$ leave behind a factor of $S_{j \ell}$. Putting this all together, and defining $\mathcal{S} = S^{\otimes n}$ as before, we see that
\begin{align}
    \dim(\Ha(g,J)) & = \sum_\ell (S_{0\ell})^{-(2g+1) + 3 - n} S_{\ell j_1} \cdots S_{\ell j_n} \\&= \sum_\ell (S_{0\ell})^\chi \bra{\ell}^{\otimes n} \mathcal{S} \ket{J} \label{eqn:dimension}
\end{align}
where $\chi = 2-2g-n$ is the Euler characteristic of the Seifert surface $\Sigma_{g,n}$. 
Therefore, plugging this expression into the link state and taking a complex conjugate because $\dim(\Ha(g,J))$ is real,
\begin{align}
    \ket{g,n,[f]=\Id} &= \sum_{J,\ell} (S_{0\ell})^\chi \mathcal{S}\ketbra{J} \mathcal{S}^\dagger\ket{\ell}^{\otimes n} 
    \\&= \sum_\ell (S_{0\ell})^\chi\ket{\ell}^{\otimes n} \label{eqn:chi_state}
    \\&\equiv \ket{\chi}
\end{align}
where we went from the first to the second line via a resolution of the identity, and used the fact that $\mathcal{S}$ is unitary. We can see immediately see that when the monodromy $[f]$ is trivial, the \emph{only} relevant data for the link state is the Euler characteristic of its Seifert surface and the number of link components. All such states are entangled in a GHZ-like manner because of the manifest form the state takes, in analogy to the three-party GHZ state $\ket{\text{GHZ}} = (\ket{000} + \ket{111}) / \sqrt{2}$. Thus, as we will see below, partial traces over any link component will lead to a separable density matrix on the remaining subsystem, and furthermore, the entanglement entropy of any subsystem is identical, independent of the particular subsystem. 

What class of links/backgrounds does trivial monodromy correspond to? By \eqref{eqn:fibration}, link complements with trivial monodromy are homeomorphic to the simple product $\Sigma_{g,n} \times S^1$. But something stronger is true: using the inverse of the argument that equated the link complements \eqref{eqn:phi_ind1} and \eqref{eqn:phi_ind2}, this will also occur whenever the link fibration takes the form $\phi(\Sigma_{g,n}) \times S^1$ for any $\phi$. This means we can take $\partial \phi(\Sigma_{g,n})$ to be any genus $g$, $n$-component link that we wish. We have just shown that for any link $\La^n$, there exists some background manifold $M$ such that its link complement $M(\La^n)$ has GHZ-like entanglement. This demonstrates that the entanglement structure of a Chern-Simons link state is not simply a function of the link, but also of the background manifold the link is embedded within. 

\subsection{How entangled is $\ket{\chi}$?} \label{sec:chient}

\begin{comment}
    
\begin{figure}
    \centering
    \includegraphics[width=0.5\linewidth]{figs/ent_triv_un.png}
    \caption{The entropy of any bipartition of boundary tori when $[f]=\Id$. The asymptotic value of each graph is $\ln(2)$.}
    \label{fig:ent_triv_un}
    \includegraphics[width=0.5\linewidth]{figs/ent_triv_norm.png}
    \caption{The same graph, but each curve is normalized by the maximum entropy $\ln(k+1)$. The large $|\chi|$ value for each graph is $\frac{\ln(2)}{\ln(k+1)}$.}
    \label{fig:ent_triv_norm}
\end{figure}
\end{comment}

\begin{figure}
    \centering
    \includegraphics[width=0.85\linewidth]{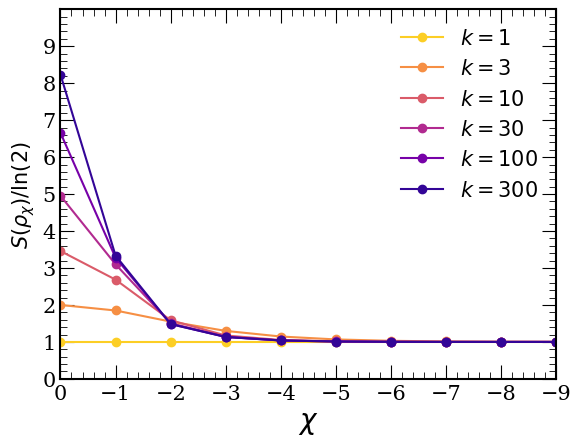}
    \caption{The entropy of any bipartition of boundary tori when $[f]=\Id$. Each plot has a maxima of $\ln(k+1)$ at $\chi=0$, and an asymptotic value of $\ln(2)$ as $\chi \to -\infty$.}
    \label{fig:ent_triv_un}
    %\includegraphics[width=0.5\linewidth]{figs/ent_triv_norm.png}
    %\caption{The same graph, but each curve is normalized by the maximum entropy $\ln(k+1)$. The large $|\chi|$ value for each graph is $\frac{\ln(2)}{\ln(k+1)}$.}
    %\label{fig:ent_triv_norm}
\end{figure}

The family of trivial-monodromy states $\ket{\chi}$ are GHZ entangled, but it is interesting to consider \emph{how} entangled they are. To analyze this numerically, we pick the gauge group $\SU(2)$ and analyze the entropy of various trivial-monodromy link states.
In Fig.~\ref{fig:ent_triv_un}, we plot the entanglement entropy across an arbitrary bipartition of link components, which turns out to be independent of said bipartition, as a function of the Euler character $\chi$ for various levels $k$. Because entanglement requires that $n \geq 2$, we only plot the cases when $\chi \leq 0$. For any level, these all reach their maxima of $\ln(k+1)$ when $\chi=0$, and levels off to a minimum of $\ln(2)$ by around $\chi \leq -6$. Thus, the entanglement entropy becomes a poor measure for distinguishing this class of states around $\chi \leq -6$.

The level independent minimum of $\ln(2)$ is easy to understand analytically. From \eqref{eqn:chi_state}, it is easy to see that for any partial trace $\rho_\chi$ of $\ket{\chi}$,
\begin{equation}
    \rho_\chi = \frac{1}{\Omega}\sum_\ell (S_{0\ell})^{2\chi}\ketbra{\ell}^{\otimes m}\,,
\end{equation}
where $\Omega = \sum_\ell (S_{0\ell})^{2\chi}$. Note that this explicitly implies that the entanglement entropy of $\ket{\chi}$ is independent of the choice of bipartition. As $\chi \to -\infty$, the relative magnitudes of the coefficients of $\rho_\chi$ decrease exponentially, explaining why the numerical value of the entropy of these states converge so quickly. In this limit, the reps $\ell$ with minimal value of $S_{0\ell}$ will dominate. The fact that the entropy minimum is $\ln(2)$ indicates that there are two $\ell$ values with minimal $S_{0\ell}$: checking numerically, these are when $\ell = 0,\frac{k}{2}$. Thus, for $|\chi|$ large enough,
\begin{align}
    \rho_{\chi \leq -6} \approx \frac{1}{2}\ketbra{0}^{\otimes m} + \frac{1}{2}\ketbra{\frac{k}{2}}^{\otimes m}\,.
\end{align}
Note that different $\chi$ correspond to link complements with different topology.  Thus, the finding that $\rho$ is almost the same for all topologies with Euler number $\chi \leq -6$ says that  link states associated to wildly different manifolds (e.g., genus $3$ or $3 \times 10^6$) can be exponentially close in the boundary Hilbert space.

\subsection{General link states and non-trivial monodromy} \label{sec:generallinkstate}

%In this section, we derive a general formula for any fibered link state, even when $[f]$ is non-trivial.In the next section, we compute link states for concrete examples, and analyze the role of the monodromy $[f]$ on the entanglement structure of $\ket{M(\La^n)}$.

We now consider the case of more general monodromy. 
For each $J$, the unitary $K_f[J]$ representing the monodromy $[f]$ in \eqref{eqn:bcycle} is finite dimensional.  Let $\{e^{i\theta_n[J]}\}$ be the eigenvalues of $K_f[J]$.  Then we define $\rho_J(\theta)$ to be the density of eigenvalue phases
%$K_f[J]$, so that if $\{e^{i\theta_n[J]}\}$ are the eigenvalues of $K_f[J]$, 
\begin{equation}
    \rho_J(\theta) = \frac{1}{\dim(\Ha(g,J))} \sum_n \delta(\theta - \theta_n[J])\,.
\end{equation}
The $\rho_J(\theta)$ is deteremined by the monodromy $[f]$, and is normalized so that for all $J$,
\begin{align}
    \int d\theta \, \rho_J(\theta) = 1 \,.
\end{align}
With this quantity in hand, we can decompose the trace $\tr(K_f[J])$ needed in \eqref{eqn:bcycle} as
\begin{equation}
    \frac{\tr(K_f[J])}{\dim(\Ha(g,J))} =  \int d\theta \, e^{i\theta} \rho_J(\theta) = \widetilde{\rho}_J(1) \,,
\end{equation}
where $\widetilde{\rho}_J(1)$ is the Fourier transform of $\rho_J(\theta)$ evaluated at $1$. We refer to $\widetilde{\rho}_J(1)$ as $\widetilde{\rho}_J$ when convenient. 
Other Fourier modes $\widetilde{\rho}_J(\alpha)$ correspond to $\tr(K_f[J]^\alpha)$, and because $K_f$ is a representation of $\text{Mod}(g,n)$, we can think of $\alpha$ as a kind of time parameter measuring how many times the monodromy $[f]$ is enacted. Note that the normalization of the eigenvalue density means that $\widetilde{\rho}_J(0) = 1$ for all $J$. Furthermore, the positivity of $\rho_J(\theta)$ implies that 
\begin{align}
    |\widetilde{\rho}_J(\alpha)| \leq |\widetilde{\rho}_J(0)| = 1 \label{eqn:positivity}
\end{align} 
for all $\alpha$ and all $J$. 

The decomposition $\tr(K_f[J]) = \widetilde{\rho}_J \dim(\Ha(g,J))$ is useful because we already computed the dimensions $\dim(\Ha(g,J))$ explicitly above. Using this substitution in \eqref{eqn:bcycle},
\begin{align}
    \ket{M(\La^n)} &= \sum_J \widetilde{\rho}_J \dim(\Ha(g,J)) \mathcal{S}\ket{J}
    \\& = \sum_{J}  \widetilde{\rho}_J \mathcal{S} \ketbra{J} \mathcal{S}^\dagger \sum_\ell (S_{0\ell})^\chi \ket{\ell}^{\otimes n}
    \\&:= \mathcal{P}(f) \ket{\chi}
\end{align}
where
\begin{align}
    \mathcal{P}(f) &= \mathcal{S} \left[ \sum_J \widetilde{\rho}_J \ketbra{J} \right] \mathcal{S}^\dagger \,, \label{eqn:monodromyop}
   \\\ket{\chi} &= \sum_\ell (S_{0\ell})^\chi\ket{\ell}^{\otimes n}\,.
\end{align}

This decomposition of $\ket{M(\La^n)}$ reveals many surprising features. First, $\ket{\chi}$ is GHZ-like for any $\chi$, and does not depend on the monodromy $[f]$ at all; the effects of the monodromy all come from the action of the operator $\mathcal{P}(f)$. Furthermore, the previous section demonstrated that $\mathcal{P}(\Id) = \Id$. Therefore, if $\ket{M(\La^n)}$ is \emph{not} GHZ-like, then it must be because of some property of the monodromy. In other words, the monodromy of a fibered link controls the entanglement structure of its associated link state.
This reduces the problem of understanding multipartite entanglement of link states in Chern-Simons theory to classifying how the topological properties of $[f]$ affect the structure of $\mathcal{P}(f)$. We call $\mathcal{P}(f)$ the monodromy operator.

$\mathcal{P}(f)$ has many properties that hold for arbitrary fibered link states. First, \eqref{eqn:monodromyop} makes it clear that $\mathcal{P}(f)$ is diagonalized by the fusion matrices $\mathcal{S}$, which in particular is a local unitary. Second,  by \eqref{eqn:positivity}, the eigenvalues $\widetilde{\rho}_J$ of $\mathcal{P}(f)$ are constrained to take values within the unit disk of the complex plane by unitarity of $K_f[J]$. $\mathcal{P}(f)$ itself, however is not always unitary.\footnote{To ensure that $\ket{M(\La^n)}$ is properly normalized, we generally need to divide by a state-dependent normalization constant.} Additionally, $\mathcal{P}(f)$ generally does not form a representation of $\text{Mod}(g,n)$.
Nevertheless, the above presentation of the state is sufficient to demonstrate that the monodromy is the link invariant which determines the entanglement structure of fibered link states. 
Finally, the eigenvalues $\widetilde{\rho}_J$ of $\mathcal{P}(f)$ are intimately tied to the topological properties of the monodromy $[f]$, as we will see below. This often makes $\mathcal{P}(f)$ an efficient way to describe the link state $\ket{M(\La^n)}$.

Anticipating the following sections, note that one could imagine two broad ways that a link state could exhibit GHZ entanglement. The first is that the monodromy operator could be a local on the disjoint link components, in which case the link state would remain GHZ-like because $\ket{\chi}$ is. Alternatively, we could imagine that the monodromy operator itself maps states onto the GHZ subspace of $\Ha(T^2)^{\otimes n}$, independently of the structure of $\ket{\chi}$. We will see examples of both phenomena shortly.

\section{Three kinds of monodromy}
\label{sec:threekinds}

%\subsection{Dehn twists generate the mapping class group}

%Computing the traces $\tr(K_f[J])$ and their associated monodromy operators $\mathcal{P}(f)$ can be complicated if the monodromy $[f]$ is complicated. 
%But formally, there is a well defined algorithm which allows us to construct $K_f[J]$ and take its trace, which we now explain. This will lead us to define the ``twisting surface'' preparing the trace $\tr(K_f[J])$, a notational device we will use often for the rest of the paper.

We now explain an algorithm for computing $K_f[J]$ and taking its trace.
As described above, $[f]$ should be thought of as a representative of the mapping class group $\text{Mod}(g,n)$, which is the group of homeomorphisms of $\Sigma_{g,n}$ that pointwise fix its boundary,
modulo isotopy. This is simply the generalization of the role $\SL(2,\Z)$ plays on the torus. In fact, $\text{Mod}(1,0) = \SL(2,\Z)$.
It is well known that $\text{Mod}(g,n)$ is finitely generated \cite{mappingclass,GERVAIS2001703,humphries} by Dehn twists around different cycles of $\Sigma_{g,n}$. This is a generalization of the fact that $\SL(2,\Z)$ is finitely generated by Dehn twists $T_1$ and $T_2=ST_1S^{-1}$ around the $a$ and $b$ cycles of the torus, respectively. Letting $T_a$ label this basis of generators for $\text{Mod}(g,n)$, this means that we can decompose $[f]$ as
\begin{align}
    [f] = T_{a_1} T_{a_2} \cdots T_{a_m}
\end{align}
for some sequence of twists of cycles labeled by $a_1, \cdots a_m$. Because $K_f$ is a representation,\footnote{Technically, it can be a projective representation, but such an overall phase will not affect the physical link state itself, so we ignore this complication.} this implies that
\begin{align}
    K_f = K_{a_1} \cdots K_{a_m}
\end{align}
where we have abbreviated $K_{T_{a}} \equiv K_a$. So if we know the action of $K_a$ on $\Ha(g,J)$ for each Dehn twist $T_a$, we can build $K_f$ for arbitrary $[f]$, and then take its trace. 

For any fixed basis, some twists have simpler representations than others. Using the basis construction outlined in Sec.~\ref{sec:pants} and Fig.~\ref{fig:pants},
%and Fig.~\ref{fig:dehn_basis}, respectively, 
twists which intersect any leg of the embedded graph in Fig.~\ref{fig:pants} transversely have a simple form.\footnote{For example, twists which traverse between neighboring holes of the surface intersect the vertical lines of the embedded graph of Fig.~\ref{fig:pants} once. Dehn twists which circle the part of the surface represented by the horizontal lines of the embedded graph also have a simple representation. An example of a twist which is more difficult to represent is one which goes ``around'' a hole of the surface. }
We can enact these twists by inserting a twisted cylinder between two punctures before they are glued together (see Fig.~\ref{fig:hopflink}). For each $j,\ell$, this can be thought of as a map on the Hilbert space $T^j_{\ell}: V^j_{\ell 0} \to V^j_{\ell 0}$, as a cylinder is just a three punctured sphere with one puncture capped off (the zero in $V^j_{\ell 0}$). Because 
\begin{align}
    \dim(V^j_{\ell 0}) = N^j_{\ell 0} = \delta^j_{\ell}
\end{align}
by the Verlinde formula \eqref{eqn:verlinde}, we see that $T^j_{\ell}$ must be zero unless $j = \ell$. If $j=\ell$, then it is a unitary acting on a one dimensional Hilbert space: a pure phase. More careful consideration reveals that \cite{Moore:1988qv,DiFrancesco:1997nk} up to a global phase proportional to the central charge,
\begin{align}
    T^j_{\ell} = e^{2\pi i h_j} \delta^j_{\ell} \,,\label{eqn:dehntwist}
\end{align}
where $h_j$ is the conformal weight of the primary operator $\phi_j$ associated to the representation $j$. Concretely, if $G_k = \SU(2)_k$, then \cite{DiFrancesco:1997nk}
\begin{align}
    h_j = \frac{j(j+1)}{k + 2} \,.
\end{align}
At the end of the day, this means in practice that if $[f]$ has a decomposition where no Dehn twists intersect, then we can find a \emph{single} basis of $\Ha(g,J)$ which diagonalizes all those Dehn twists, making their trace particularly simple to compute. To compute the trace of $K_f$ in this case, we simply insert factors of $T^j_{\ell}$ representing the corresponding $K_a$ between the fusion matrices arising from the pair of pants Hilbert spaces that split along that cycle, and sum the repeated indices. 

The representations of intersecting Dehn twists are more complicated. The reason is that for each non-commuting twist, one must use duality transformations to change from the starting basis to a more convenient one, find the action in a simple form, and invert those duality transformations again. The procedure is outlined explicitly in \cite{Moore:1988qv}, and in practice means that one must compute $\tr(K_f[J])$ numerically.  In this paper we will evaluate examples that do not require this more elaborate procedure, although there is no obstacle to implementing it.

\paragraph{The twisting surface:} In general, we can efficiently represent the monodromy by drawing an ordered sequence of curves $a_1, \cdots a_m$ on $\Sigma_{g,J}$, which represent the Dehn twists around these curves. We refer to this presentation as the ``twisting surface'' for preparing $\tr(K_f[J])$. It is sometimes convenient to think of the twisting surface as representing the state $\ket{M(\La^n)}$ itself, as a kind of two-dimensional path integral presentation of the state. See Fig.~\ref{fig:hopflink}, Fig.~\ref{fig:k4} or Fig.~\ref{fig:hnpants} for examples of the twisting surface for the Hopf link, Hopf keyring, or Hopf chain, respectively.

\subsection{The Nielsen-Thurston classification}

A major benefit of presenting the link state $\ket{M(\La^n)}$ in terms of the data $(g,n,[f])$ is that mathematicians have classified how the properties of the monodromy $[f]$ relate to the topological classification of links. The Nielsen-Thurston classification \cite{mappingclass,NTclass} proves that there are three types of monodromy: 

\begin{enumerate}
    \item $[f]$ is periodic, i.e., $[f]^N$ is isotopic to the identity for some $N \in \N$, up to boundary-isotopic Dehn twists.\footnote{Boundary-isotopic Dehn twists are Dehn twists around curves which surround a single puncture, and are equivalent to  local unitaries acting on the link state.} Furthermore, we require that $[f]$ is not reducible (see definition below).
    Torus links in $S^3$, namely links that can be drawn on the surface a torus within $S^3$ are an example. In the case of knots on $S^3$, \emph{only} torus knots (torus links with one component) have periodic monodromy \cite{miyazaki}.  
    \item $[f]$ is pseudo-Anosov, which for our purposes just means that $[f]$ is not of the other two types. A link has pseudo-Anosov monodromy if and only if its link complement is hyperbolic, i.e., $M(\La^n)$ can be endowed with a complete hyperbolic metric. This is a deep theorem due to Thurston \cite{mappingclass,NTclass}. This is the ``generic'' case.
    \item $[f]$ is reducible, which means there is a system of disjoint, non-boundary isotopic, closed curves $$\gamma_1 \cup \cdots \cup \gamma_m := \gamma \subset \Sigma_S$$ such that $f(\gamma) = \gamma$. We usually take $\gamma$ to be the maximal such multicurve. The Hopf keyring and Hopf chain (see the next section) are examples, as are all satellite knots. A reducible monodromy can be further analyzed by cutting $\Sigma_S$ along $\gamma$, where $[f]$ reduces piecewise to one of the other two cases on each component of $\Sigma_S \setminus \gamma$.
\end{enumerate}

The pseudo-Anosov case is the most difficult to analyze directly, and we will leave it to future work. In the next section, we will explain the consequences of periodic monodromy on the entanglement structure of fibered link states. But first, using the machinery we have developed so far, we compute link states in some examples. This will allow us to discuss in more detail the role of the monodromy $[f]$ in determining  entanglement structure.

\subsection{Examples}

In the examples below we take the background manifold $M$ to be the three sphere $S^3$, unless otherwise noted.

\subsubsection{The Hopf Link} \label{sec:hopflink}

\begin{figure}
    \centering
    \includegraphics{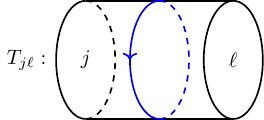}

    \caption{The twisting surface for the Hopf link. This twisting surface is the same as the cylinder preparing $T_{j\ell}$ that is inserted along various cycles in more general cases. %\VB{fix boundary not puncture}
    }
    \label{fig:hopflink}
\end{figure}

The Hopf link ($2^2_1$ in Rolfsen notation) has a Seifert surface homeomorphic to the punctured 2-sphere $\Sigma_{0,2}$ (equivalently, a cylinder), and monodromy $[f] = T^{-1}$ for our chosen handedness of the link components, which consists of a single Dehn twist around the cylinder (Fig.~\ref{fig:hopflink}).
This is exactly the twisting surface which prepares $T^j_{\ell}$. Therefore, taking the two punctures to have representations\footnote{The conjugation of $j$ to $\bar{j}$ accounts for the fact that one representation must be viewed as ``inflowing'' and the other must be viewed as ``outflowing'' because $T^j_\ell$ has an upper and lower index.} $J = \bar{j}\ell$, 
\begin{align}
    \tr(K_{f}[\bar{j}\ell]) = [T^j_{\ell}]^* = e^{-2\pi i h_j} \delta^j_{ \ell}
\end{align}
Inserting this in \eqref{eqn:bcycle}, and repeatedly using resolutions of the identity, we can show that
\begin{align}
    \ket{2^2_1} &= \sum_{j \ell} [T^j_{\ell}]^* \mathcal{S} \ket{\bar{j}}\ket{\ell}
    \\&= \sum_{j \ell ab} e^{-2\pi i h_j} \delta^j_{ \ell} S_{a\bar{j}} S_{b\ell} \ket{a}\ket{b}
    \\&= \sum_{ab} (S T^\dagger S^\dagger )_{ba}\ket{a}\ket{b}
    \\&= \Id \otimes (ST^\dagger S^\dagger )\left[\sum_\ell \ket{\ell}\ket{\ell}\right]
\end{align}
where $\sum_\ell \ket{\ell}\ket{\ell}$ is the $\ket{\chi=0}$ state from \eqref{eqn:chi_state}.   Normalizing the state, we have determined that
\begin{align}
    \ket{2^2_1} &= \mathcal{P}(f) \ket{\chi = 0} \\
    \mathcal{P}(f) &= \mathcal{S} \left[\Id \otimes T^\dagger \right] \mathcal{S}^\dagger \\
    \ket{\chi =0} &= \frac{1}{||\chi||}\sum_\ell (S_{0\ell})^{\chi = 0} \ket{\ell}^{\otimes 2}
\end{align}
where $\chi = 0$ is the Euler characteristic of the Seifert surface $\Sigma_{0,2}$, $||\chi||$ is a normalization constant, and $\ket{\chi=0}$ is the state of a link with the same Seifert surface but trivial monodromy. The non-trivial monodromy is captured by the action of $\mathcal{P}(f)$; we explained the reason for this above.
One way to see why $\mathcal{P}(f)$ is a local unitary in this case is as follows. 
We can act any local unitary on $\ket{M(\La^n)}$ by gluing a collar defining that unitary onto each puncture. Because such a procedure completely defines the monodromy in this case, we know in advance that it will act on $\ket{\chi=0}$ via something like $\mathcal{P}(f)$.

It was noted in \cite{Balasubramanian_2017} that the Hopf link acts as the maximally entangled state, i.e., like a Bell pair, if one studies its entanglement structure. We have reproduced this result. However, 
%noting that $S = S^\dagger$ for $\SU(2)_k$, 
the authors of \cite{Balasubramanian_2017} found that the link state  took the form
\begin{align}
    \ket{2^2_1, \text{previous}} = \Id \otimes S\ket{\chi=0}
\end{align}
which differs with our results by a local unitary transformation. What is the source of this difference?

The answer is we have chosen a different framing of the Hopf link (see \cite{Witten:1988hf} for an explanation of framing). To see this, we change the framing of each component of our Hopf link by $-1$ units. %Geometrically, this ``undoes'' the natural framing induced by the link components twisting around each other on $\Sigma_S$. 
Quantum mechanically, this has the effect of acting the local unitary $T^\dagger \otimes T^\dagger$ on our link state:
\begin{align}
    T^\dagger \otimes T^\dagger\ket{2^2_1} &= (T^\dagger \otimes T^\dagger)(\Id \otimes ST^\dagger S^\dagger ) \ket{\chi=0}
    \\&= T^\dagger \otimes (T^\dagger ST^\dagger S^\dagger ) \ket{\chi=0}
\end{align}
But $\ket{\chi=0}$ is the maximally entangled two component state, so we can use the transpose map (and the fact that $T^\dagger$ is symmetric) to rewrite this as
\begin{align}
    T^\dagger \otimes T^\dagger\ket{2^2_1} &= \Id  \otimes (T^\dagger ST^\dagger S^\dagger T^\dagger) \ket{\chi=0}\,.
\end{align} 
One can use $\SL(2,\Z)$ commutation relations to show that $T^\dagger ST^\dagger S^\dagger T^\dagger = S$. Thus, we find that 
\begin{align}
    T^\dagger \otimes T^\dagger\ket{2^2_1} &= \Id \otimes S \ket{\chi=0}
    \\&= \ket{2^2_1, \text{previous}} \,,
\end{align}
which now \emph{exactly} reproduces the result in \cite{Balasubramanian_2017}. What led to this difference of framing? The natural choice of framing in their case was the ``zero framing'' because of the way they constructed the link state. For our methods, the natural choice of framing was the one induced by the surface $\Sigma_{g,n}$. This is related to the map $\mathcal{U}_\phi$ defined in the previous section. In general, changes of framing will only change our states by at most a local unitary. As this will not affect the entanglement structure of our link states, we will not discuss framing in the remainder of this paper.

\subsubsection{The Hopf Keyring}

\begin{figure}
    \centering
    \includegraphics[]{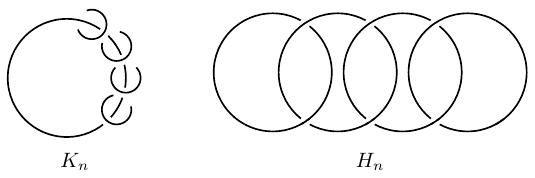}

    \caption{The family of Hopf keyrings ($K_n$) and the Hopf chain ($H_n$).}
    \label{fig:knhn}
\end{figure}

\begin{figure}
    \centering
    \includegraphics[]{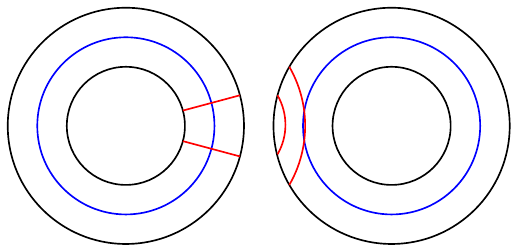}
    
    \includegraphics[]{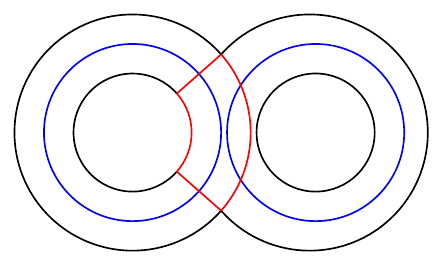}
    
    \caption{Hopf plumbing $K_3 = H_3$. Top: Each annulus (concentric black rings)  is homeomorphic to the Seifert surface of a Hopf link. The monodromy is represented by a right-handed twist around the blue circles. The black circles are the link components, which become the $J$-punctures after gluing in the defects. The red arcs are included to guide the eye in visualizing the gluing.  Bottom: The result of Hopf plumbing. }
    \label{fig:hopfplumb}
\end{figure}

\begin{figure}
    \centering
    \includegraphics[]{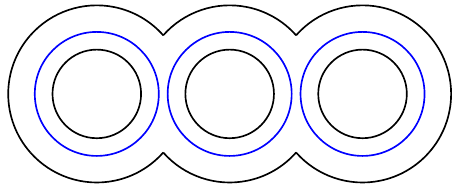}

    \caption{To build the twisting surface for $K_4$, we plumb a third Hopf link in the same manner shown in Fig.~\ref{fig:hopfplumb}.}
    \label{fig:k4}
\end{figure}

The $n-$Hopf keyring $K_n$ is an infinite class of links that we draw in Fig.~\ref{fig:knhn}. Using the technique of Hopf plumbing \cite{HARER1982263}, we can build the twisting surface for $K_n$ by appropriately gluing together the Seifert surfaces of $n-1$ Hopf links. Hopf plumbing involves gluing the Seifert surface of the Hopf link to another Seifert surface in the manner indicated in Fig~\ref{fig:hopfplumb}. It is a special case of an operation known as a Murasugi sum \cite{murasugi}. The important fact about Hopf plumbing is that the resulting link is still fibered, with Seifert surface given by the Hopf plumbing procedure. This is demonstrated for $K_3$ in Fig.~\ref{fig:hopfplumb} and $K_4$ in Fig.~\ref{fig:k4}. Because we can cut the monodromy along the blue curves in Fig~\ref{fig:hopfplumb}, $K_n$ is a reducible link for any $n$. The monodromy can also be computed explicitly: it is simply the monodromy of each Hopf link concatenated together. Because each Dehn twist acts locally on the correspnding puncture, from the twisting surface perspective we can ``pinch off'' each of the Dehn twists, as they will act on $\ket{K_n}$ via local unitaries, analogously to the Hopf link. Thus, we can think of the Hopf keyrings as links with trivial monodromy times a local unitary.  We then immediately know that
\begin{align}
    \ket{K_n} &= \mathcal{P}(f) \ket{\chi = 2-n} \\
    \mathcal{P}(f) &= \mathcal{S} \left[T \otimes \cdots \otimes T \otimes \Id \right] \mathcal{S}^\dagger \\
    \ket{\chi = 2-n} &= \frac{1}{||\chi||}\sum_\ell (S_{0\ell})^{2-n} \ket{\ell}^{\otimes n}
\end{align}
where $||\chi||$ is a normalization constant, and $\ket{\chi = 2-n}$ is GHZ-like, which we demonstrated in Sec.~\ref{sec:trivial}. The structure of this state is explained in Sec.~\ref{sec:generallinkstate}. %The $\Id$ in $\mathcal{P}(f)$ acts on the main component which links with all the others. 
In this case, $\mathcal{P}(f)$ is a local unitary, and so $\ket{K_n}$ and $\ket{2-n}$ have the same entanglement structure. Thus,  $\ket{K_n}$ is GHZ-like for all $n$, and the entropy is numerically captured by the discussion in Sec.~\ref{sec:chient}.
%Once again, we note that the general structure we derived in Sec.~\ref{sec:generallinkstate} is again demonstrated in this infinite family of examples. 
This again reproduces (and generalizes) the results of \cite{Balasubramanian_2018}, who demonstrated the GHZ nature of $K_3$, up to a choice of framing.

\subsubsection{The Hopf Chain} \label{sec:hn}

\begin{figure*}
    \centering
    \includegraphics[]{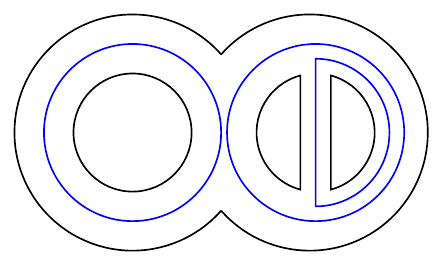}
    \includegraphics[]{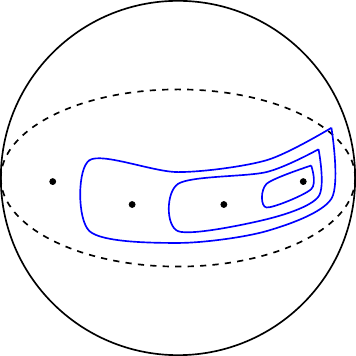}
    \caption{To build the twisting surface for $H_4$, we plumb another Hopf link into $H_3$ as shown above (recall that $H_3 = K_3$). Left: The twisting surface of $H_4$. Note the difference with $K_4$. Right: The same twisting surface as at left, but with  boundaries shrunk to punctures (black dots). }
    \label{fig:h4}
\end{figure*}

\begin{figure*}
    \centering
    \includegraphics[]{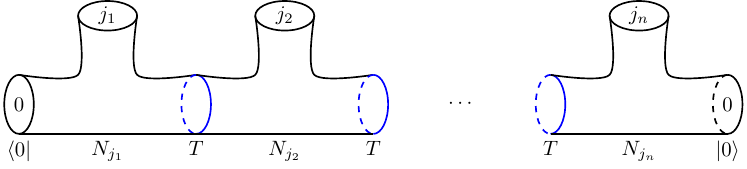}
    \caption{The twisting surface for $H_n$}
    \label{fig:hnpants}
\end{figure*}

A  similar, but subtly different, family of links is the Hopf chain $H_n$ (Fig.~\ref{fig:knhn}). This is also defined by plumbing $n-1$ Hopf links together, but in a different pattern than the Hopf keyrings. Both links have genus zero (their Seifert surfaces are homemorphic to punctured spheres as shown in Figs.~\ref{fig:k4} and \ref{fig:h4}) and $n$ components, so the only difference is their monodromy. Similarly to $K_n$, the $H_n$ are reducible links because we can cut the surface into distinct pieces, for example along the blue curves in Fig.~\ref{fig:h4} and \ref{fig:hnpants}. We draw the twisting surface for $H_4$ in Fig.~\ref{fig:h4} and for $H_n$ in Fig.~\ref{fig:hnpants}.  We will see that the Hopf chains are examples of links with monodromy which is not reducible to local unitaries. We will now compute the link state $\ket{H_n}$.

Using the pair-of-pants decomposition suggested in Fig.~\ref{fig:hnpants}, we see that there is a recursive structure to this monodromy. A simple application of the rules described above for computing the trace of the monodromy shows that
\begin{align}
    \tr(K_f[J]) = \sum_{\ell_i,\ell_i'} N^0_{ j_1 \ell_1} T^{\ell_1}_{ \ell'_1} N^{\ell_1'}_{ j_2 \ell_2} T^{\ell_2}_{ \ell'_2} \cdots  N^{\ell_{n-1}'}_{ j_n 0} \,.
\end{align}
To get this expression, we could construct $K_f[J]$ by creating the tensor product of the identity on each three-punctured sphere Hilbert space $V_{ik}^j$ with the appropriate $T_{j\ell}$ cylinders inserted along the blue curves, and taking the trace of the resulting operator. But an easier way to construct the same trace is to simply use the twisting surface of Fig.~\ref{fig:hnpants} to read off the trace immediately. From now on, we will focus on using the twisting surface to directly compute the traces.
We will now do some algebra to simplify this expression, but the reader may wish to skip to Eq.~\ref{eqn:I2}. 

\paragraph{Simplifying $\tr(K_f[J])$:} Using the Verlinde formula Eq.~\ref{eqn:verlinde}, %and the Dehn twist formula Eq.~\ref{eqn:dehntwist},

\begin{align}
    \tr(K_f[J]) &= \sum_{\alpha_i}  \frac{S_{j_1 \alpha_1}}{S_{0\alpha_1}} (S^\dagger T S)_{\alpha_1 \alpha_2} \cdots (S^\dagger T S)_{\alpha_{n-1} \alpha_n}\frac{S_{j_n \alpha_n}}{S_{0\alpha_n}}(S^\dagger)_{\alpha_n 0} S_{0\alpha_1}\label{eqn:hn_monodromy_components}
    \\&=\sum_{\alpha_i} \bra{J} \mathcal{S} \ket{\vec{\alpha}} \frac{\bra{\vec{\alpha}} \mathcal{S}^\dagger [ \mathcal{T} \otimes \ketbra{0}] \mathcal{S} \sigma_n \ket{\vec{\alpha}}}{\bra{0^{\otimes n}} \mathcal{S}\ket{\vec{\alpha}}}
\end{align}

where $\mathcal{T} = T^{\otimes n - 1}$, and $\sigma_n$ is the cyclic permutation operator $$\sigma_n\ket{\alpha_1, \cdots, \alpha_{n}} = \ket{\alpha_2, \cdots, \alpha_1}\,.$$ 
It is convenient to replace the dummy indices $\vec{\alpha}$ with their dual representations, which does not change the sum because it is a dummy index. Then, using the $S$ transformation identities \eqref{eqn:Ssym}, 
\begin{align}
    \tr(K_f[J])=\sum_{\alpha_i} \bra{J} \mathcal{S}^\dagger \ket{\vec{\alpha}} \frac{\bra{\vec{\alpha}} \mathcal{S} [ \mathcal{T} \otimes \ketbra{0}] \mathcal{S}^\dagger \sigma_n \ket{\vec{\alpha}}}{\bra{0^{\otimes n}} \mathcal{S}\ket{\vec{\alpha}}}\,.
\end{align}

Therefore, using Eq.~\ref{eqn:bcycle} and a resolution of the identity for $\ket{J}$, the link state is
\begin{equation}
    \ket{H_n} =\sum_{\alpha_i} \frac{\bra{\vec{\alpha}} \mathcal{S} \, [ \mathcal{T} \otimes \ketbra{0}] \,  \mathcal{S}^\dagger \sigma_n \ket{\vec{\alpha}}}{\bra{0^{\otimes n}} \mathcal{S} \ket{\vec{\alpha}}} \ket{\vec{\alpha}}\,.
\end{equation}
For the moment, focus on the numerator.
Use the $\SL(2,\Z)$ identity $S T S^\dagger = T^\dagger S^\dagger T^\dagger$ 
%$\mathcal{S}^\dagger \mathcal{T} \mathcal{S} = \mathcal{T} \mathcal{S}^\dagger \mathcal{T}$ %and that $\ket{0}^{\otimes n} = \sigma_n \ket{0}^{\otimes n}$ 
to see that
\begin{equation}
     \bra{\vec{\alpha}} \mathcal{S} \, [ \mathcal{T} \otimes \ketbra{0}] \,  \mathcal{S}^\dagger \sigma_n \ket{\vec{\alpha}} = \bra{\vec{\alpha}} (\mathcal{T}^\dagger \otimes S )( \mathcal{S}^\dagger\otimes  \ketbra{0} ) ( \mathcal{T}^\dagger \otimes S^\dagger) \sigma_n \ket{\vec{\alpha}} \,.
\end{equation}
Because $T^\dagger\ket{\alpha} = e^{-2\pi i h_\alpha}\ket{\alpha}$ and $(\mathcal{T}^\dagger \otimes \Id) \sigma_n = \sigma_n (\Id \otimes \mathcal{T}^\dagger)$, 

\begin{equation}
   \bra{\vec{\alpha}} (\mathcal{T}^\dagger \otimes S )( \mathcal{S}^\dagger\otimes  \ketbra{0} ) ( \mathcal{T}^\dagger \otimes S^\dagger) \sigma_n \ket{\vec{\alpha}}
    = e^{2\pi i ( h_{\alpha_1} - h_{\alpha_n})}
    \bra{\vec{\alpha}} [\mathcal{S}^\dagger \otimes  (S \ketbra{0} S^\dagger )]\sigma_n \ket{\vec{\alpha}}  \,.
\end{equation}

Returning to the full expression for $\ket{H_n}$, we can reabsorb these phases by acting $\mathcal{T}$ on the vector $\ket{\vec{\alpha}}$ itself:
\begin{equation}
    \ket{H_n} =\sum_{\alpha_i} \frac{\bra{\vec{\alpha}} (\mathcal{S}^\dagger\otimes  (S \ketbra{0} S^\dagger ))\sigma_n \ket{\vec{\alpha}} }{\bra{0^{\otimes n}} \mathcal{S}\ket{\vec{\alpha}}} (T_1 \otimes \Id \otimes T^\dagger_n)\ket{\vec{\alpha}}\,.
\end{equation}
This is just a change of framing, and in particular is a local unitary. For notational convenience, we neglect this factor of $T^2$ for the remainder of this section.

\paragraph{The simplified link state:}
We can rearrange the resulting formula to a more enlightening form by defining the map
\begin{align}
    I_2 &= \sum_\alpha  (S_{0\alpha})^{-1} \ket{\alpha}^{\otimes 2} \bra{\alpha} S^\dagger\,,\label{eqn:I2}
    \\ I_2 & : \Ha(T^2)^{\otimes m} \to \Ha(T^2)^{\otimes m + 1} \,.
\end{align}
We think of $I_2$ as acting on the first tensor factor of any Hilbert space $\Ha(T^2)^{\otimes m}$ it acts on, and the identity elsewhere. Similarly defined maps will appear again  later, and one can trace their origin to the appearance of the fusion matrices $N^j_{ik}$ in the calculation of the monodromy $\tr(K_f[J])$. 
After some algebra, we can rearrange the link state to take the form
\begin{align}
    \ket{H_n} =[\bra{0} S^\dagger]_1 \underbrace{I_2 \cdots I_2}_{n} S \ket{0} \,.\label{eqn:hnnesting}
\end{align}
In this expression, each $I_2$ is a map from $\Ha(T^2)^{\otimes m} \to \Ha(T^2)^{\otimes m+1}$, and entangles the new copy of $\Ha(T^2)$ with the previously added component. Furthermore, the projection $[\bra{0}S^\dagger]_1$ contracts on the first tensor factor of the resulting $\Ha(T^2)^{\otimes n+1}$. Thus, the string of maps in Eq.~\ref{eqn:hnnesting} produces a state in $\Ha(T^2)^{\otimes n}$, as expected.
The fact that $\ket{H_n}$ can be represented in this way shows that it is a ``matrix product state'' \cite{Cirac_2021}, with matrices $I_2$ of dimension $k+1$, which in particular, do not depend on $n$. 
Intuitively, this presentation of $\ket{H_n}$ shows that it is produced by a nested sequence of bipartite entangling operations, knitting together a more intricate multipartite structure. 

Although the steps to obtain Eq.~\ref{eqn:hnnesting} took some algebra, there is a simple, physical way to understand this decomposition. When we add a new component to $H_n$ to create $H_{n+1}$, one way to describe this process is as follows. Take the unentangled pair $(0_1,H_n)$ and entangle the unknot $0_1$ with the first component of $H_n$ in order to link them: this is exactly the role of $I_2$.

\begin{figure}
    \centering
    \includegraphics[width=0.95\linewidth]{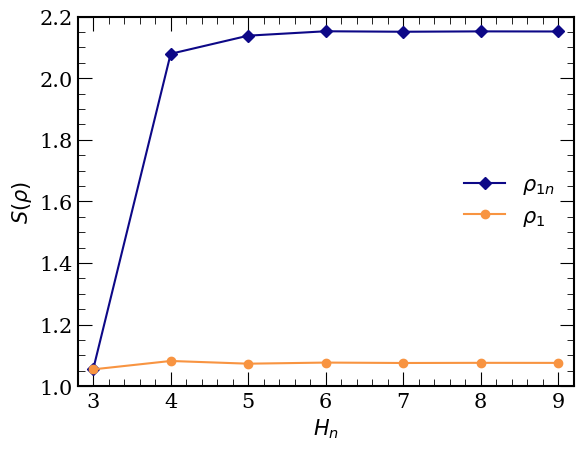}
    \caption{The entropy of both one end of $H_n$ ($\rho_1$) and both ends together $(\rho_{1n}$) for $n=3,4,5,6$ at level $k=4$. The fact that these disagree for $n>3$ proves that $\rho_{1n}$ is not separable, and therefore those link states are not GHZ entangled. The fact that they do agree when $n=3$ is a nice check that $\ket{H_3}$ displays GHZ entanglement.}
    \label{fig:hnents}
\end{figure}

We will now show that $\ket{H_n}$ is $W$-like for $n \geq 4$ using the following diagnostic. If $\ket{\Psi} = \sum_\ell \sqrt{p_\ell} \ket{\ell}^{\otimes n}$ is a GHZ-like state, then if we trace out \emph{any} $m \geq 1$ components of $\ket{\Psi}$ to get the reduced state $$\rho_m = \sum_\ell p_\ell \ketbra{\ell}^{\otimes n-m}\,,$$ then the entropy $S(\rho_m)$ is independent of $m$. Thus, the entropy of GHZ states is independent of the subsystem we choose to analyze. In particular, this means that if the entropy of two subsystems of a link state disagree, the state must contain W-like entanglement. We can numerically confirm that $\ket{H_n}$ contains $W$-like entanglement if we choose $n=4$ and $G = \SU(2)$ for various levels by computing the entanglement entropy for various bipartitions of tori, which we present in Fig.~\ref{fig:h4Wlike}. As these do not agree for every bipartition (and is not bipartite entangled), the state is W-like entangled. 
%Thus, $H_n$ is W-like when $n \geq 4$. 
By contrast, when $n=3$, there are only two monodromy curves on the twisting surface, one which encircles two punctures, and another which encircles a single puncture. But the first curve can be deformed to only encircle the previously unenclosed puncture. So precisely when $n=3$, the monodromy is secretly just a local unitary acting on the trivial twisting surface of $\chi=-1$. This explains geometrically why the special case of $H_3$ has GHZ entanglement --  this is the only case where the twisting surface can be homeomorphed into such a form. For all larger $n$, no such trivializing-deformation is possible, so all other $H_n$ will be W-like. We confirm this numerically for $n=4,5,6$ and level $k=4$ in Fig.~\ref{fig:hnents}.

\begin{figure}
    \centering
     \begin{tabular}{|c|c|c|}
    \hline
        Level & $S(\rho_{14})$ & $S(\rho_1)$ \\
        \hline
        \hline
        1 & 1.38629436 & 0.69314718 \\
        2 & 2.07944154 & 1.08219553 \\
        3 & 2.62166284 & 1.34897237 \\
        4 & 2.96996892 & 1.55036031 \\
        5 & 3.2986487 & 1.71133755 \\
        6 & 3.53634631 & 1.84500026 \\
        7 & 3.75584901 & 1.95903219 \\
        8 & 3.93914771 & 2.0583098 \\
        9 & 4.10981582 & 2.14611336 \\
        10 & 4.2476002 & 2.2247508 \\
        \hline
    \end{tabular}
    \caption{
    %\VB{placeholder with the data, also I forgot $\rho_{14}$.} 
    The entropy of various bipartitions of tori for $\ket{H_4}$. Note that $\rho_1 = \tr_4 \rho_{14}$, and so the fact the entropy changes means that there is residual entanglement between links $1$ and $4$: the state is not seperable, so $\ket{H_4}$ is not GHZ-like. 
    }
    \label{fig:h4Wlike}
\end{figure}

\subsection{Lessons learned}\label{sec:lessons}

We found that the Hopf keyring states $\ket{K_n}$ produced a separable state after tracing out an arbitrary subset of link components. This implies that $\ket{K_n}$ are GHZ entangled for all $n$. We can see the GHZ nature of the Hopf keyring geometrically from a path integral perspective. $\ketbra{K_n}$ is prepared with two copies of the twisting surface with reversed orientations. If we bipartiton the tori $J \mapsto (J_a,J_b)$, then the partial trace 
\begin{equation}
    \rho_a = \frac{1}{\braket{K_n}}\tr(\ketbra{K_n})_{J_b}
\end{equation}
is computed by gluing the $J_b$ punctures on the two copies together. Because of the CPT theorem, any local unitaries around each puncture will always cancel when glued together. 
Equivalently, the monodromy operator $\mathcal{P}(f)$ is a local unitary in this case, and will not affect the entanglement entropy.
Then because the Euler characteristic of two-surfaces is additive under gluing, regardless of the bipartition of $ (J_a,J_b)$ the final twisting surface computing $\tr(\rho_a^n)$ will always be the same -- the closed surface of Euler characteristic $2n\cdot \chi$, with all monodromy canceled out. Thus, the R\'enyi entropy $$S_n(\rho_a) = \frac{1}{1-n} \ln \tr(\rho_a^n)$$ is independent of the bipartition of links. In the limit $n \to 1$, the R\'enyi entropy becomes the von Neumann entropy, so the von Neumann entropy will also be independent of the bipartition of link components.
This is a consequence of GHZ-like entanglement, and the same argument will hold for any link with trivial monodromy (up to local unitaries).

On the other hand, we found that the Hopf chain states $\ket{H_n}$ did not produce a separable state if $n \geq 4$. This W-like entanglement is again easy to see geometrically from the perspective of the twisting surface. The intricate nesting of monodromy, shown in Fig.~\ref{fig:hnpants}, means that the replica twisting surface preparing $\tr(\rho_a^n)$ \emph{will} generically depend on the gluing procedure. In turn, this means that the R\'enyi entropies will depend on the choice of bipartition of boundary tori, and so the state must be W-like entangled. This was numerically confirmed for $H_n$ for $n=4,5,6$ in Fig.~\ref{fig:h4Wlike} and Fig.~\ref{fig:hnents}.

\section{Periodic monodromy}
\label{sec:periodic}

In the previous section, we gave examples in which GHZ-like entanglement arose from the action of a local monodromy operator on a GHZ-like state associated to link complements with trivial monodromy.  In this section we will show another way in which GHZ-like entanglement can arise, namely if the monodromy is periodic.  We will show this in general, after discussing the example of torus links in $S^3$.

%Based on the previous section, a natural question is if a local monodromy operator is the \emph{only} way GHZ entanglement could arise. The answer is no, as we will see in this section. Before we discuss the general case of periodic monodromy, it is useful to examine the classic example of such a link: torus links on $S^3$. 

%This section is structured as follows. 
Up to this point, our main strategy has been to canonically quantize the Chern-Simons path integral along the $S^1$ of the fibration in \eqref{eqn:fibration}. The reason this was convenient is because this $S^1$ always exists for any fibered link. But for torus links, there is another slicing of the path integral that introduces a different $S^1$ we can canonically quantize along. We will find that this alternate slicing gives a clearer picture for why torus links have GHZ entanglement. We will then explain why an analogous ``alternate'' slicing always exists for links with periodic monodromy. This will show that periodic monodromy implies a GHZ-like entanglement structure.

%First, as a warm-up, we recall the nice description of a torus link via its braid definition, which allows for a (relatively) simple computation of the link state $\ket{T(p,q)}$. We can use this braid description to determine the trace of the monodromy $\tr(K_f[J])$ by a simple change of basis of $\ket{T(p,q)}$ (outlined in \eqref{eqn:jonesmonodromy}). This will give us insight into the interesting topological structure behind the monodromy of torus links. Next, we confirm these interesting features directly from the path integral, explaining the necessary mathematics along the way. Finally, we explain why these features hold for \emph{any} link with periodic monodromy, not just torus links. 

\subsection{Torus links via orbifold quantization} 

Any link in $S^3$ which can be drawn on the surface of a torus is called a \emph{torus link}. These links  are completely specified by a pair of integers $(p,q)$, measuring how many times each link component wraps the torus along each cycle. In terms of its braid closure,  we can think about torus links as being represented by the braid $B(p,q) = (\sigma_1 \cdots \sigma_{p-1})^q$.\footnote{A braid can be thought of as a series of vertical strands which can swap places any number of times, with a choice of which strand passes over the other. $\sigma_i$ represents a particular transposition of the $i$th strand crossing over the $i+1$th strand. A braid closure results from identifying the first strand at the bottom with the first strand on the top, and so on. See \cite{Witten:1988hf} for more details.}
A torus link has $n = \gcd(p,q)$ components, each of which is a $(p/n,q/n)$ torus knot. The monodromy of a torus link is periodic (up to local unitaries), and is known for any $(p,q)$. It is derived e.g. in \cite{bell2012monodromieshomogeneouslinks}, and can be written as a product of $(p-1)(q-1)$ Dehn twists. We could use the methods we have developed so far to compute the  link state, but because these twists intersect each other, finding their exact representation on the Seifert surface Hilbert space is complicated. However, it  turns out to be helpful to slice the path integral in a different way. For more general links with periodic monodromy, an analogous alternate slicing will always be possible.

Because torus links have a convenient braid representation, we will expand the link state $\ket{T(p,q)}$ in the $\ket{J}$ basis, as opposed to the $\mathcal{S}\ket{J} $ basis as we have done in the rest of this paper. In other words, we want to use the Chern-Simons path integral to compute the colored Jones polynomial
\begin{align}
    V_J(p,q) = \braket{T(p,q)}{J} \,.
\end{align}

Of course, this result is well known in the literature, and is explained in e.g. in \cite{Balasubramanian_2018,Isidro_1993,Labastida_2000,Brini_2012}. From these references, the colored Jones polynomial for a $(p,q)$ torus link (in our notation) takes the form
\begin{align}
    V_{J}(p,q) = \sum_{\ell} \dim(\Sigma_{0,J + \ell}) V_\ell(p/n,q/n)\,. \label{eqn:torusjones}
\end{align}
Here, $V_\ell(p/n,q/n)$ is the colored Jones polynomial of the $(p/n,q/n)$ torus \emph{knot}.
Based on our discussion in Sec.~\ref{sec:trivial}, we can already see that torus links have GHZ monodromy because of the factor of $\dim(\Sigma_{0,J + \ell})$ in the sum, in analogy to the factor of $\dim(\Ha(g,J))$ in \eqref{eqn:triviallinkstate}. We are going to compute $V_J(p,q)$ from scratch because the new procedure we will introduce gives insight into why this factor occurs, and why this feature is shared by other periodic links.

To begin, arrange the Wilson lines to lie within a single solid torus of $S^3$. This works for any link, and is one way to identify a link with its braid closure. Because $S^3$ is the union of two such tori, we can therefore think of $V_J(p,q)$ as
\begin{align}
    V_J(p,q) = \bra{0} S \ket{B_J(p,q)}\,.
\end{align}
$\ket{B_J(p,q)}$ is the state on $\Ha(T^2)$ which is prepared by the braid closure of Wilson lines in one solid torus, and $\ket{0}$ is the state on $\Ha(T^2)$ with no Wilson lines inserted. They are glued together with a modular $S$ transformation so the resulting background manifold is $S^3$. We can insert a resolution of the identity to see that 
\begin{align}
    V_J(p,q) = \sum_\alpha S_{0}^{\alpha} \braket{\alpha}{B_J(p,q)}\,.
\end{align}

Here, $\alpha$ is a representation of $G$ at level $k$, so $\ket{\alpha}$ is a state in $\Ha(T^2)$. The term $\braket{\alpha}{B_J(p,q)}$ in this sum is sometimes called the $\alpha$-th braid trace of $B_J(p,q)$. In terms of the Chern-Simons path integral, $\braket{\alpha}{B_J(p,q)}$ can be prepared as follows. Take the sphere $S^2$ and puncture it $p+1$ times. We think of $p$ of these punctures as being arranged evenly around the equator of $S^2$, and the remaining puncture as sitting at, say, the north pole. Cyclically assign the $p$ punctures along the equator representations $j_1, \cdots j_n$, so there are $\frac{p}{n}$ copies of each representation. Call this collection of representations $J_p$. Furthermore, we put the representation $\bar{\alpha}$ at the north pole, conjugated because it came from a bra.
Multiply this punctured sphere by an interval, and identify the endpoints of this interval according to the map specified by the particular braid $B(p,q)$. In this case, the braid $B(p,q) = (\sigma_1 \cdots \sigma_{p-1})^q$ acts as a $2\pi/p$ rotation around the $\bar{\alpha}$ puncture, applied $q$ times. 
Based on the above description, we can see that 
\begin{align}
    \braket{\alpha}{B_J(p,q)} = Z[\Sigma_{0,J_p+ \alpha}\times_{B(p,q)} S^1]_\alpha\,.
\end{align}
We use the $\alpha$ subscript to emphasize the representation of the extra puncture. It is a lower index because the puncture has a representation $\bar{\alpha}$. Topologically, the Wilson lines in this path integral really do form the $(p,q)$ torus link.

This is very similar to the setups we have considered up to this point --  quantizing along the $S^1$ of the path integral, the braid $B(p,q)$ plays essentially the same role that the monodromy $[f]$ played earlier in this work. But $B(p,q)$ is not the same thing as the monodromy of the torus link introduced in previous sections. They are conceptually distinct, and are in some sense $S$-dual to each other, though not in a direct way. Regardless, using the same techniques, we can treat the $\frac{q}{p}$ twist around the $\bar{\alpha}$ puncture as a local unitary and think of it as being applied to the untwisted base $\Sigma_{0,J_p+ \alpha}$. This local unitary has matrix elements
\begin{align}
    (T^{q/p})^\beta_{\alpha} = e^{2\pi i (q/p) h_\beta} \delta^\beta_\alpha \,.
\end{align}
We do this to simplify the path integral calculation of the braid trace, which now takes the form
\begin{align}
    \braket{\alpha}{B_J(p,q)} = \sum_{\beta,\gamma}  Z[\Sigma_{0,J_p+ \gamma} \times S^1]_\gamma \delta^{\gamma}_{\beta}(T^{q/p})^\beta_{\alpha} 
    \,.
\end{align}
We include the factor of $\delta^\gamma_\beta$ to emphasize that the path integrals preparing $Z[\Sigma_{0,J_p+ \alpha} \times S^1]_\gamma$ and $(T^{q/p})^\beta_{\alpha} $ are to be glued using the identity mapping.

Because the punctured sphere $\Sigma_{0,J_p+ \gamma}$ is smooth (up to defects), we could quantize in the usual way along this $S^1$, and would obtain the correct answer after using some $S$ matrix identities. However, it is helpful to first notice that $\Sigma_{0,J_p+\gamma}$ has a $\Z_{p/n}$ rotational symmetry around the $\bar{\alpha}$ puncture because the representations of each puncture are repeated. Thus, we can quotient the sphere by this discrete symmetry and treat $\Sigma_{J_p + 1} / \Z_{p/n}$ as an orbifold. This orbifold is just the $n+1$ punctured sphere, now with all distinct representations at the punctures. The path integral on this space is just the dimension of the punctured sphere Hilbert space $\dim(\Sigma_{0,J+\gamma})$. However, the gluing map between the $\gamma$ puncture the twist $(T^{q/p})^\beta_{\alpha}$ will be affected by the orbifold procedure. 
This change in gluing  means that the $\delta^\gamma_\beta$ is replaced by some other linear operator, which we call $C(p/n)^\gamma_\beta$. This is the same map $C(p/n)^\gamma_\beta$ that appears in \cite{Balasubramanian_2018}. Note that the quotiented path integral and unquotiented path integral do not agree unless this gluing map is introduced. In other words,
\begin{align}
    Z[\Sigma_{0,J_p + \gamma} \times S^1]_\gamma \neq Z[\Sigma_{0,J_ n+ \gamma} \times S^1]_\gamma\,,
\end{align}
but 
\begin{align}
    Z[\Sigma_{0,J_p + \beta} \times S^1]_\beta= Z[\Sigma_{0,J_ n+ \gamma} \times S^1]_\gamma C(p/n)^\gamma_\beta\,.
\end{align}
The fact that these path integrals agree is easiest to see from the three dimensional perspective: taking the braid closure of $p$ strands wrapping $q$ times is the same as the braid closure of taking $n = \gcd(p,q)$ strands wrapping $q/p$ times. The benefit of beginning with $p$ strands, rather than $n$, is that it makes the origin of the map $C(p/n)^\gamma_\beta$ more transparent.
Changing the gluing map is what ensures that the total path integral over the orbifold agrees with the unorbifolded one we started with.

Putting the pieces together, 
\begin{align}
    \braket{\alpha}{B_J(p,q)} &= \sum_{\beta,\gamma}   \dim(\Sigma_{0,J+\gamma}) C(p/n)^{\gamma}_{\beta}(T^{q/p})^\beta_{\alpha} \,,
\end{align}
and therefore,
\begin{align}
    V_J(p,q) = \sum_{\gamma} \dim(\Sigma_{0,J+\gamma}) \left(\sum_{\alpha,\beta} C(p/n)^{\gamma}_{\beta} S_{0}^\alpha (T^{q/p})^\beta_{\alpha} \right)  \,.
\end{align}
But Eq.~(4.10) of \cite{Balasubramanian_2018} shows that
\begin{align}
    V_\ell(p/n,q/n) =  \sum_{\alpha,\beta} C(p/n)^{\ell}_{\beta} S_{0}^\alpha (T^{q/p})^\beta_{\alpha}\,,
\end{align}
and therefore we have exactly reproduced \eqref{eqn:torusjones}. For later comparison, it is useful to rewrite this result in a slightly different notation. If we define the operator
\begin{align}
    \mathcal{O}^\dagger= S C(p/n) (T^{q/p}) S^\dagger\,,
\end{align}
we find that that torus link  states take the form
\begin{align}
    \ket{T(p,q)} = \sum_\ell (S_{0\ell})^{2-n} \left[\frac{\bra{0} \mathcal{O} S\ket{\ell}}{\bra{0}S\ket{\ell}}\right] \mathcal{S}\ket{\ell}^{\otimes n}\,.
\end{align}
We can think of $\mathcal{O}$ as the gluing operator between the orbifold circle bundle with trivial monodromy and the solid torus prepared by the trivial representation $\ket{j=0}$ we are gluing in along the singular point of the orbifold. 
%representing the monodromy around the singular point of the orbifold, because the $\gamma$ puncture is a fixed point of the $\Z_{p/n}$ quotient.

From this calculation, we have learned that the Chern-Simons path integral preparing $\ket{T(p,q)}$ can be thought of as a circle fibration with an orbifold as a base. The path integral this orbifold was easily computed by performing surgery to excise the singular point (which had representation $\bar{\alpha}$ before the orbifolding), and gluing it back in in an appropriate way with the map $\mathcal{O}$. This ``orbifold fibration'' is precisely the structure we will exploit for more general periodic links.

\subsubsection{The monodromy of torus links}

Before we discuss this orbifold procedure in more detail, in this section we will examine the monodromy of torus links in more detail.

Let $\Sigma(p,q)$ be the Seifert surface for the torus link $T(p,q)$. In \cite{misev2016plumbing}, it was shown that $\Sigma(p,q)$ can be described as a thickened version of the completely connected bipartite graph $K_{p,q}$, with each edge thickened with the blackboard framing.\footnote{The blackboard framing is the one induced by thickening the edges  of $K_{p,q}$ along the plane of the blackboard on which it is drawn.} $\Sigma(p,q)$ has genus \cite{graf2014teteatetegraphstwists} 
\begin{align}
    g = \frac{1}{2}((p-1)(q-1) + 1 - n) \,,
\end{align}
which means that a $(p,q)$ torus link has Euler characteristic 
\begin{align}
    \chi = 1- (p-1)(q-1)\,.
\end{align}
As a quick check of this formula, we note that $p=q=2$ (the Hopf link), then these formulas give the same answers as above. 

The strategy here will be to plug in \eqref{eqn:torusjones} into \eqref{eqn:jonesmonodromy} to compute the link monodromy. Using the Verlinde formula to compute the dimensions, 

\begin{align}
    \tr(K_f[J]) &= \sum_{L,\ell,\alpha} (S_{0\ell})^{1-n} \bra{\ell^{\otimes n}} \mathcal{S}\ketbra{L}\mathcal{S}^\dagger\ketbra{J}{\alpha}S^\dagger\ket{\ell} V_\alpha(p/n,q/n)^*
    \\&= \sum_{\ell,\alpha} (S_{0\ell})^{1-n} \braket{\ell^{\otimes n}}{J} \bra{\alpha}S^\dagger\ket{\ell} V_\alpha(p/n,q/n)^*\,. \label{eqn:Inpredef}
\end{align}

We have just derived the interesting feature of torus link monodromy that leads to GHZ entanglement -- the factor of $\braket{\ell^{\otimes n}}{J}$ in the sum. This factor implies that
\begin{equation}
   \tr(K_f[J]) = 0 ~ \text{unless} ~ j_1 = \cdots = j_n = \ell \,. \label{eqn:samereps}
\end{equation}
for some $\ell$. To see why this implies GHZ entanglement, it is convenient to define the family of maps $I_n:\Ha(T^2) \to \Ha(T^2)^{\otimes n}$
\begin{align}
    I_{n} = \sum_\ell (S_{0\ell})^{1-n} \ket{\ell}^{\otimes n} \bra{\ell} S^\dagger \,,\label{eqn:In}
\end{align}
which are similar in spirit to the maps defined in \eqref{eqn:I2}. While $I_2$ was a map from e.g. $\Ha(T^2) \to \Ha(T^2)^{\otimes 2}$, $I_n$ is a map from $\Ha(T^2) \to \Ha(T^2)^{\otimes n}$.  Both of these maps also have the same origin: the fusion matrices present in the decomposition of the monodromy $\tr(K_f[J])$. Crucially, $I_n$ has support on the subspace (up to local unitaries) of $\Ha(T^2)^{\otimes n}$ with GHZ entanglement, because the resulting states will be linear combinations of $\ket{\ell}^{\otimes n}$. Using this map, we see from combining \eqref{eqn:Inpredef} and \eqref{eqn:In} that the trace of the monodromy takes the form 
\begin{align}
     \tr(K_f[J]) &= \sum_{\alpha} V_\alpha(p/n,q/n)^* \bra{J} I_n \ket{\alpha}  \,.
\end{align}
We can simplify this even further by noting the torus knot $T(p/n,q/n)$ has a link state
\begin{align}
    \ket{T(p/n,q/n)} &= \sum_\alpha V_{\alpha}(p/n,q/n)^* \ket{\alpha} \,,
    \\&= \sum_\alpha \tr(K_{f'}[\alpha]) S\ket{\alpha}\,.
\end{align}
Here, $f'$ is the monodromy of the $T(p/n,q/n)$ torus knot.
%, and $g' = \frac{(p-n)(q-n)}{2n^2}$ is the genus of the torus knot Seifert surface. 
Plugging this expression into the previous equation, we see that
\begin{align}
     \tr(K_f[J]) &= \bra{J}I_n\ket{T(p/n,q/n)} \,.
\end{align}
Finally, plugging this monodromy into \eqref{eqn:bcycle} and using a resolution of the identity,
\begin{align}
    \ket{T(p,q)} &= \mathcal{S} I_n \ket{T(p/n,q/n)}
    \\&= \mathcal{S} I_n \sum_\ell \tr(K_{f'}[\ell]) S\ket{\ell} 
    \\&= \sum_\ell (S_{0\ell})^{1-n}\tr(K_{f'}[\ell])
    \mathcal{S} \ket{\ell}^{\otimes n}\,.
\end{align}

The above calculation of the link state of the $(p,q)$ torus link was also computed in \cite{Balasubramanian_2018}. We can see that the quantity they define as $f_\ell(p,q)$ now has a clearer geometric interpretation: it is the trace of the monodromy which goes around the torus knot $(p/n,q/n)$ that remains after fusing the $n$ link components together into a single torus knot. 

What have we learned from this calculation? The most important result was that any torus link state $\ket{T(p,q)}$ lies in the image of the map $I_n$, up to the local unitary $\mathcal{S}$. The image of $I_n$ is manifestly GHZ-like, so torus links have GHZ entanglement. In the language of the monodromy operator, \eqref{eqn:samereps} implies that
\begin{equation}
    \mathcal{P}(f) = \mathcal{S}\left[\sum_\ell \widetilde{\rho}_\ell \ketbra{\ell}^{\otimes n} \right]\mathcal{S}^\dagger\,. \label{eqn:GHZPf}
\end{equation}
Thus, we can see that for torus links, the monodromy operator is not a local unitary. This is because the eigenvalues do not factorize. But the image of $\mathcal{P}(f)$ is \emph{always} the GHZ subspace of $\Ha(T^2)^{\otimes n}$. The results of the next section will imply that up to local unitaries conjugating $\mathcal{P}(f)$, this remains true for any periodic $[f]$.

%This result still holds even if we replaced $\mathcal{S}$ with a more general local unitary $\mathcal{O}$. The map $I_n$ arose because of the fusion matrices in \eqref{eqn:torusjones}, so it will be crucial for us to understand why an analogous property holds for more general periodic links. For torus links, this property is most easily explained by expanding the path integral in the $a$-cycle basis. We will now examine this 

\subsection{Seifert manifolds}\label{sec:seifertmanifold}

Let $\Sigma_{g,n,m}$ be the genus $g$, $n$ punctured orbifold with $m$ isolated singular points. As long as $\Sigma_{g,0,m}$ is compact, $m$ must be finite, as the singular points are isolated. Of course, the structure of the orbifold may differ near each singular point (e.g., their conical defects may differ), so the notation $\Sigma_{g,n,m}$ is not sufficient to classify which orbifold we are discussing. So our first step will be to use describe the better notation for discussing the two dimensional orbifolds of interest to us.

To understand the structure of $\Sigma_{g,n,m}$, we instead consider the trivial circle bundle $\Sigma_{g,n+m} \times S^1$. This is a bundle we have extensively described in Sec.~\ref{sec:trivial}. If we give $n$ punctures representations $J$, and $m$ punctures the representations $L$, the Chern-Simons path integral on this bundle is
\begin{align}
    Z[\Sigma_{g,J + L} \times S^1] = \sum_{\ell} (S_{0\ell})^{2-2g-n-m} \bra{J}\mathcal{S}\ket{\ell}^{\otimes n}\bra{L}\mathcal{S}\ket{\ell}^{\otimes m} \,.
\end{align}

We now consider the effect of gluing (empty) solid tori along the $m$ punctures in the $L$ representations, in the same way that one glues the $J$ Wilson lines along the $n$ boundary tori. Let $\mathcal{O}^{(1)}, \cdots, \mathcal{O}^{(m)}$ be the operators which represent  this gluing onto each puncture. These maps are in a representation of $\text{Mod}(1,0) \sim \SL(2,\Z)$, and are determined by the coprime integers $(\alpha,\beta)$ corresponding to the relative slope between the $b$ cycles of the boundary tori. For example, if $(\alpha,\beta) = (0,-1)$, then $\mathcal{O}$ sends one $b$ cycle to the other $-a$ cycle, and so $\mathcal{O} \sim S$. The $a$ cycle of the solid torus we are gluing is contractible,\footnote{The $a$-cycle is the contractible cycle in this case because we have adopted the convention of gluing the $m$ solid tori along \emph{defects}, as opposed to boundary components. This difference in convention leads to an additional factor of $S$, which makes the $a$-cycle the contractible one.} so the other two integers $(\gamma,\delta)$ of the $\SL(2,\Z)$ transformation that $\mathcal{O}$ represents are irrelevant, as long as $\alpha \delta - \beta \gamma = 1$. At the level of operators, this is because we only don't actually care about the entire operator $\mathcal{O}$, but only its image $\bra{0}\mathcal{O}$, the actual solid torus we are gluing in. If the torus was not empty, then $(\gamma,\delta)$ would have a more direct role to play.

We are interested in the class of three-manifolds which can arise after gluing in these solid tori to trivial monodromy bundles. Such manifolds are called \emph{Seifert fibered} manifolds. This is an unfortunate terminology, because it has nothing to do with the fibration of a link complement by its Seifert surface. To avoid confusion as much as possible, we will just refer to this class of manifolds as Seifert manifolds instead. Based on the procedure defined above, we can see that Seifert manifolds are completely characterized \cite{orlik,10.1007/BF02398271} by their \emph{Seifert symbol}
\begin{align}
    S = (g,n ; (\alpha_1,\beta_1), \cdots, (\alpha_m,\beta_m))\,.
\end{align}
Note that manifolds with different Seifert symbols may be homeomorphic, though they can be ``normalized'' in such a way that this kind of degeneracy can not occur. We do not need such a normalization for our purposes, so we will not discuss this point further.

It is shown in \cite{thurston1998hyperbolicstructures3manifoldsii} that any link complement with periodic monodromy $[f]$ has a presentation as a Seifert manifold. In fact, a link complement is a Seifert manifold if and only if its monodromy is periodic, or reducible with only periodic subpieces. The boundary tori will be contained in some of these subpieces, but others will have purely bulk contributions. This is called the JSJ decomposition \cite{jaco1979seifert,Johannson1979HomotopyEO}. The orbifold base $\Sigma_{g,n,m}$ of this Seifert manifold is not the same as the Seifert surface $\Sigma_S$ of the link. For example, for torus links, we saw that orbifold base of the Seifert manifold preparing the Jones polynomial had genus $0$, but Seifert surface of torus links have genus $g = \frac{1}{2}((p-1)(q-1) + 1 - n)$, which is non-zero.\footnote{The only two exceptions are the unknot and the Hopf link.} Nevertheless, the monodromy $[f]$ of the link is what determines if the link complement is a Seifert manifold or not.

Let $\ket{S}$ be the state on $\Ha(T^2)^{\otimes n}$ prepared by the Chern-Simons path integral on a Seifert manifold with Seifert symbol $S$. Based on the above discussion, this path integral is simply 

\begin{align}
    \braket{J}{S} & = \sum_{\ell} (S_{0\ell})^{2-2g-n-m} \bra{J}\mathcal{S}\ket{\ell}^{\otimes n}\bra{0}^{\otimes m} \mathcal{O}^{(1)}\otimes \cdots \otimes \mathcal{O}^{(m)}\mathcal{S}\ket{\ell}^{\otimes m}\,,
\end{align}

where $\mathcal{O}^{(i)}$ is the operator enacting the $(\alpha_i,\beta_i)$ gluing. It is helpful to think of the $\mathcal{O}^{(i)}$ as acting on the left, changing the particular gluing of the empty solid torus we attach to the $m$ punctures.
This expression may seem complicated, but note that if we define
\begin{equation}
     \lambda_\ell \equiv  (S_{0\ell})^{2-2g-n-m} \bra{0}^{\otimes m} \mathcal{O}^{(1)}\otimes \cdots \otimes \mathcal{O}^{(m)}\mathcal{S}\ket{\ell}^{\otimes m}\,, \label{eqn:GHZspectra}
\end{equation}
then this amplitude simplifies to 
\begin{align}
    \braket{J}{S} & = \sum_{\ell} \lambda_\ell \bra{J}\mathcal{S}\ket{\ell}^{\otimes n}
\end{align}
In other words, a Seifert manifold always has a link state
\begin{align}
    \ket{S} &= \sum_{J,\ell} (\lambda_\ell \bra{J}\mathcal{S}\ket{\ell}^{\otimes n})\ket{J}
    \\&= \sum_{\ell} \lambda_\ell \,\mathcal{S}\ket{\ell}^{\otimes n} \,. \label{eqn:periodicGHZstate}
\end{align}

Thus, we have shown that link complements who are Seifert manifolds always lead to GHZ-like entanglement. 
We call the tuple of complex numbers $\lambda_\ell$ the GHZ-spectrum of $\ket{S}$.
Similar to the dependence of link states on $\phi$ explained in \eqref{eqn:phirelation}, this equation should be understood as holding up to local unitaries which control how the basis $\ket{J}$ was defined in the first place.

\subsection{Does the converse hold?} \label{sec:numerics}

\begin{figure*}
    \centering
    \includegraphics[width=0.32\linewidth]{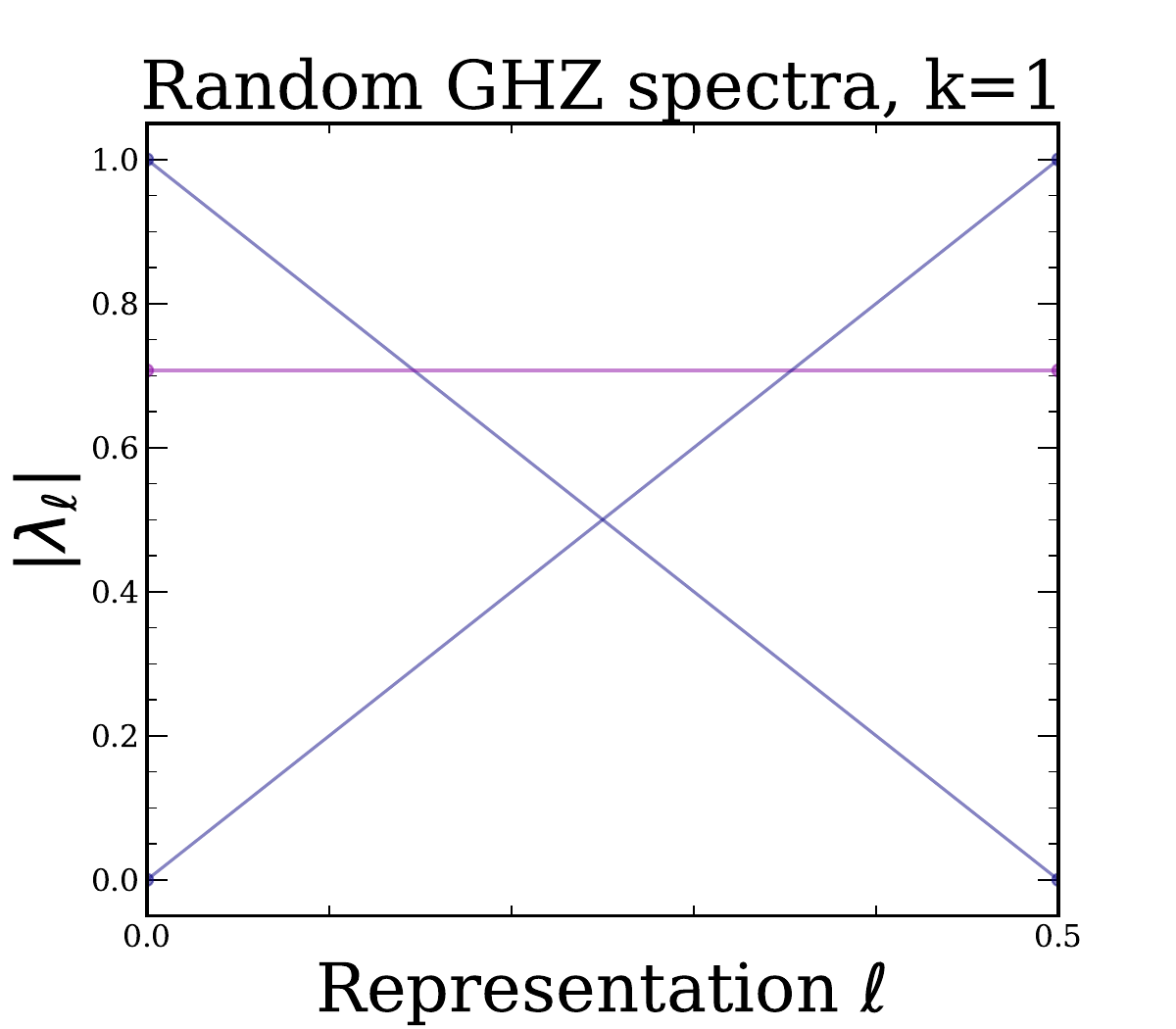}
    \includegraphics[width=0.32\linewidth]{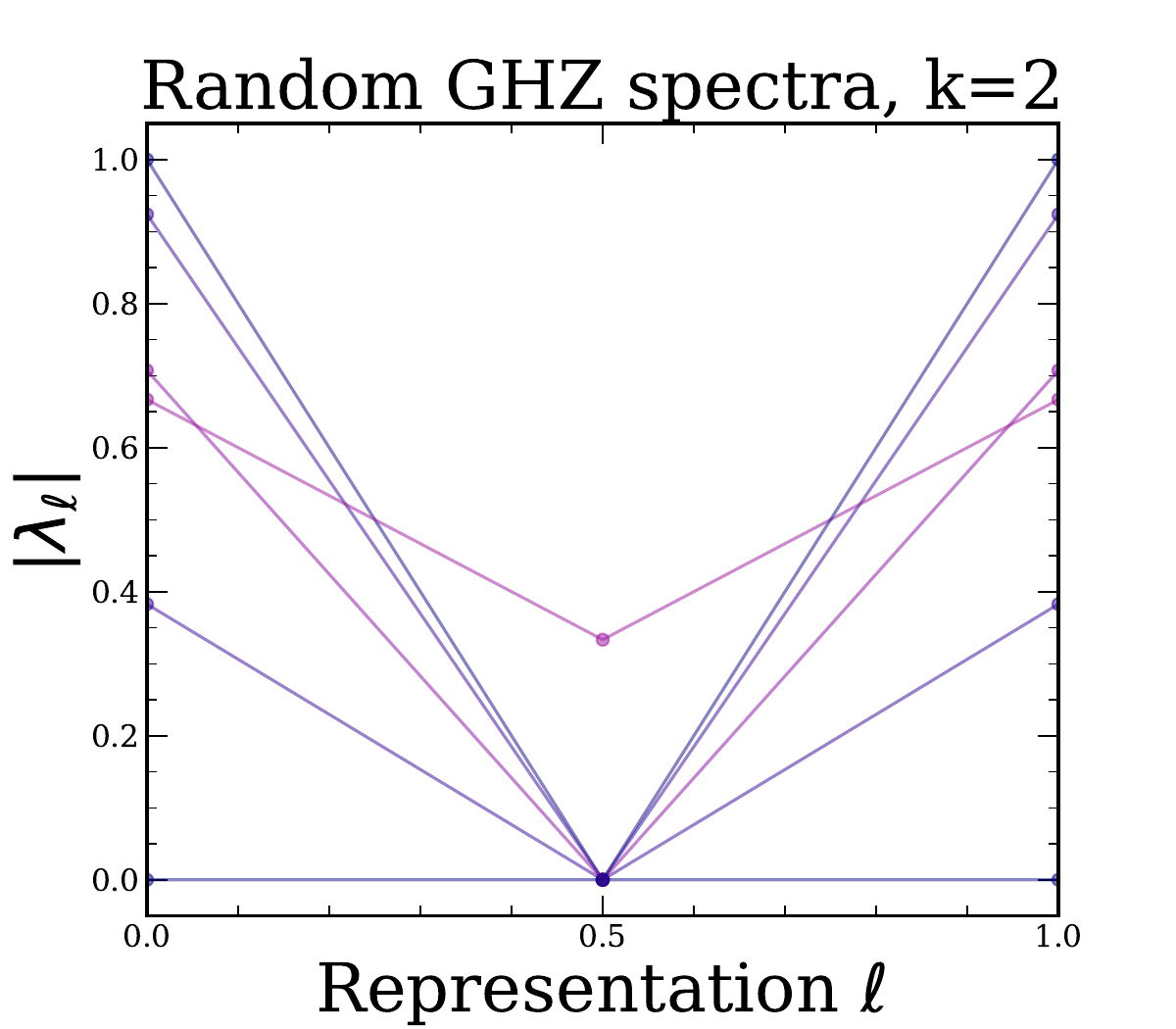}
    \includegraphics[width=0.32\linewidth]{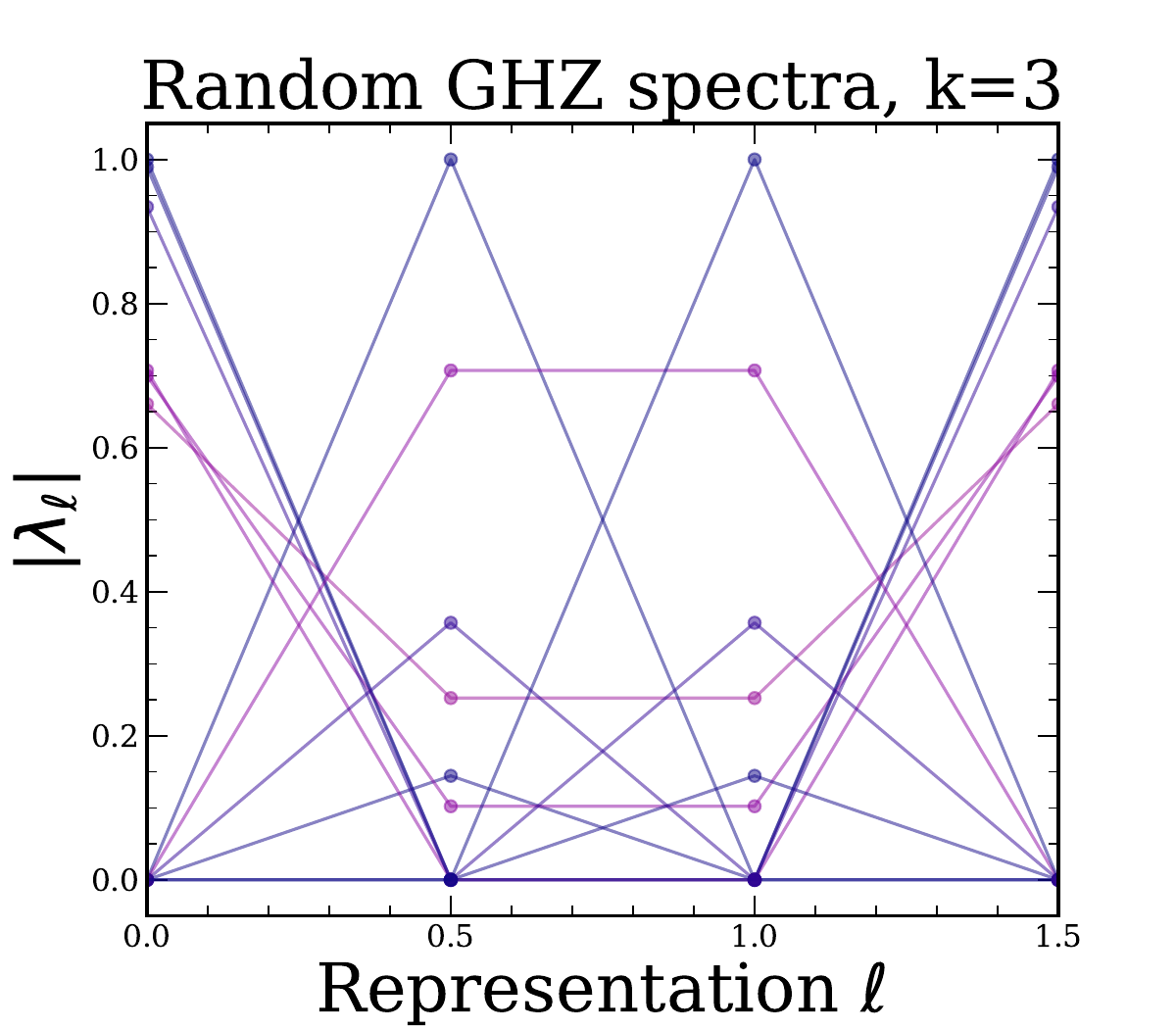}
    \includegraphics[width=0.32\linewidth]{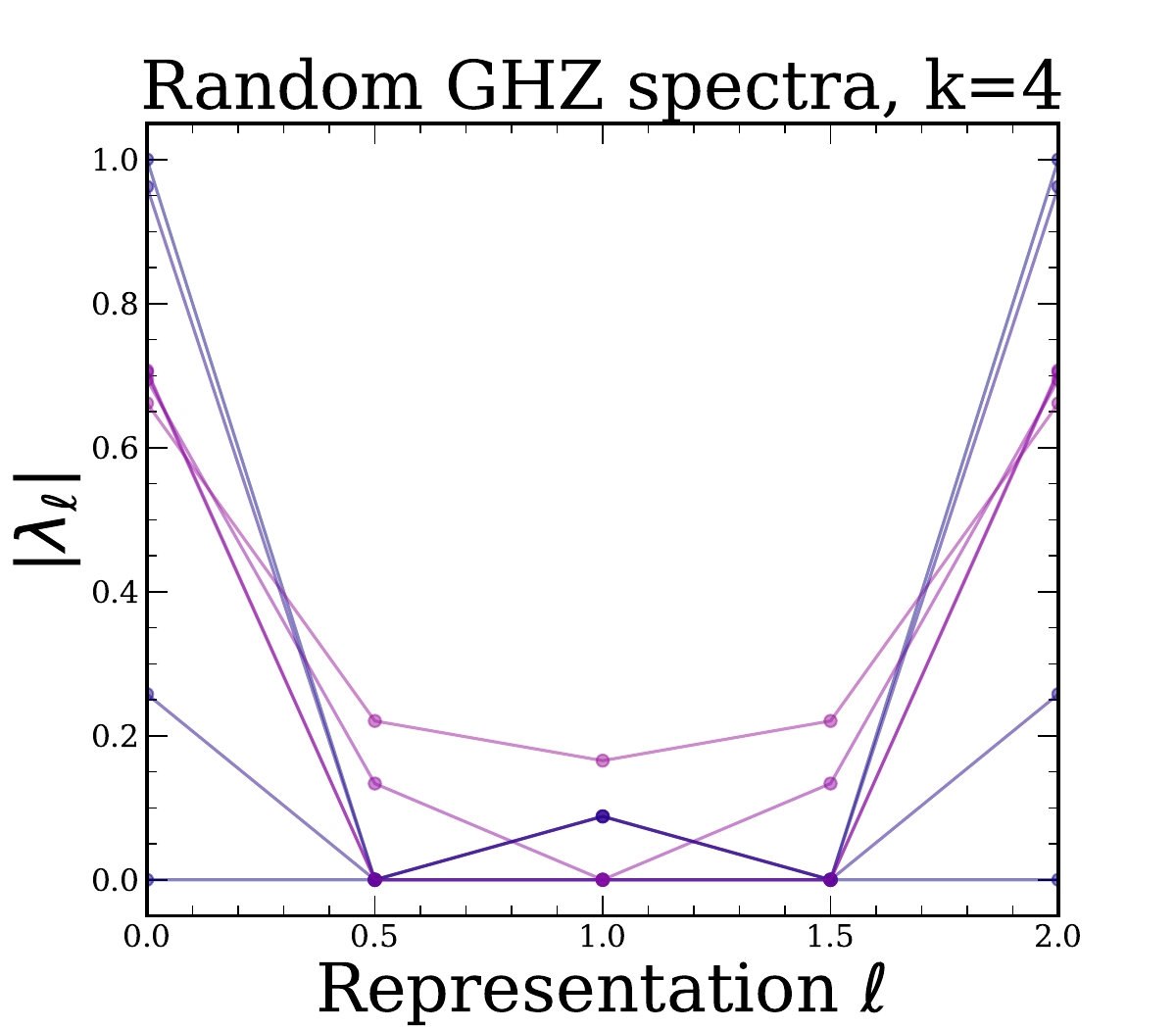}
    \includegraphics[width=0.32\linewidth]{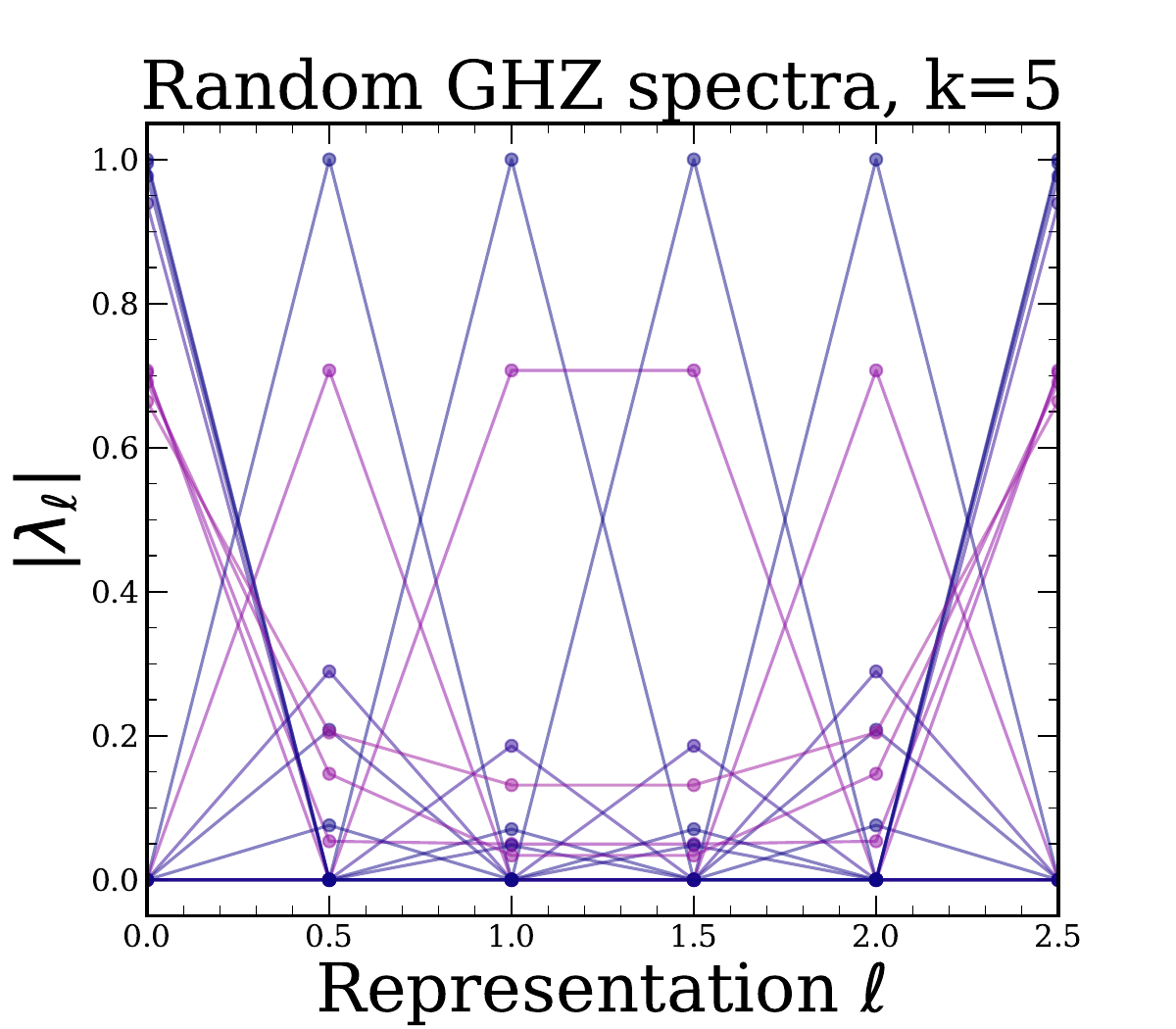}
    \includegraphics[width=0.32\linewidth]{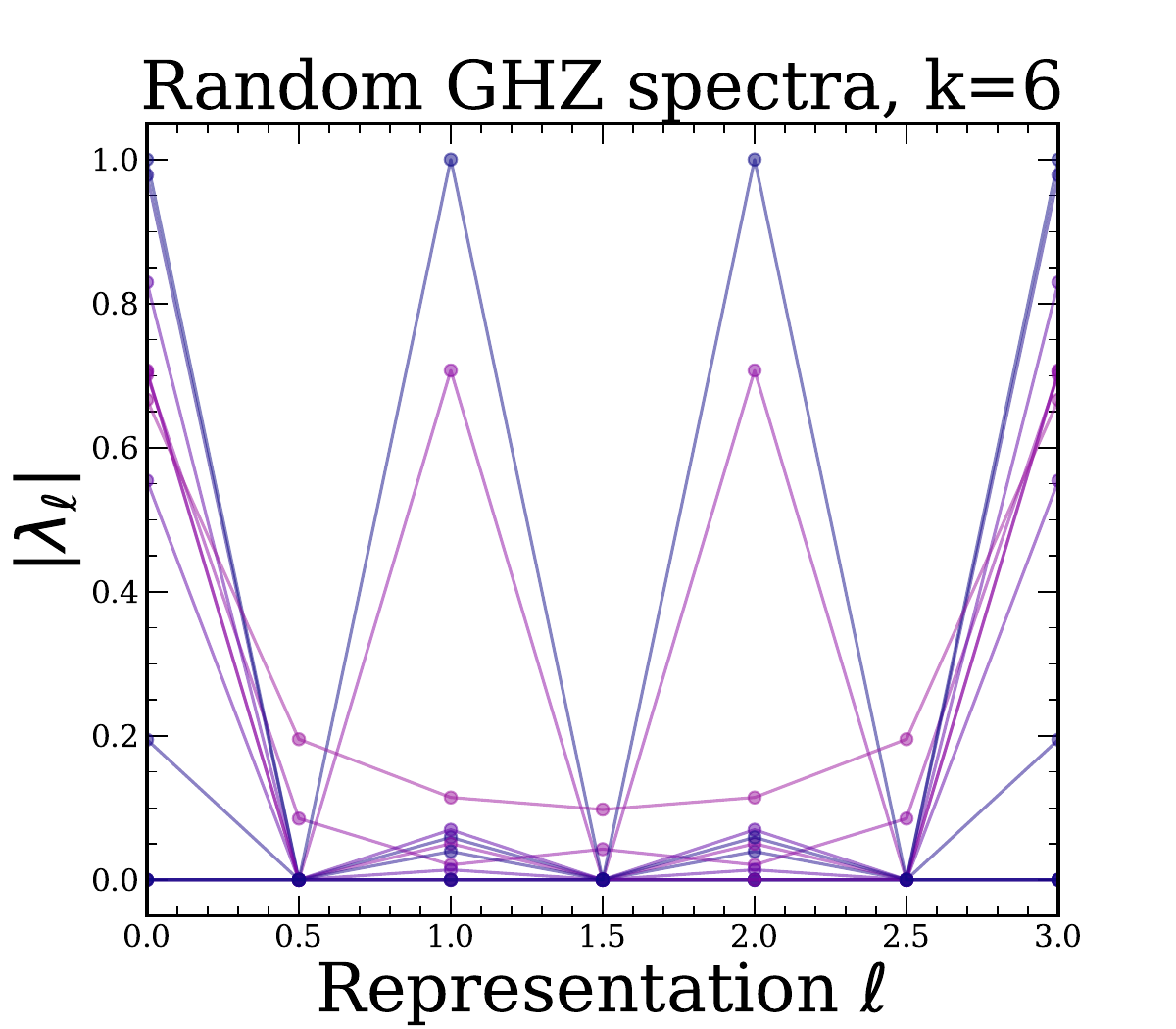}
    \includegraphics[width=0.32\linewidth]{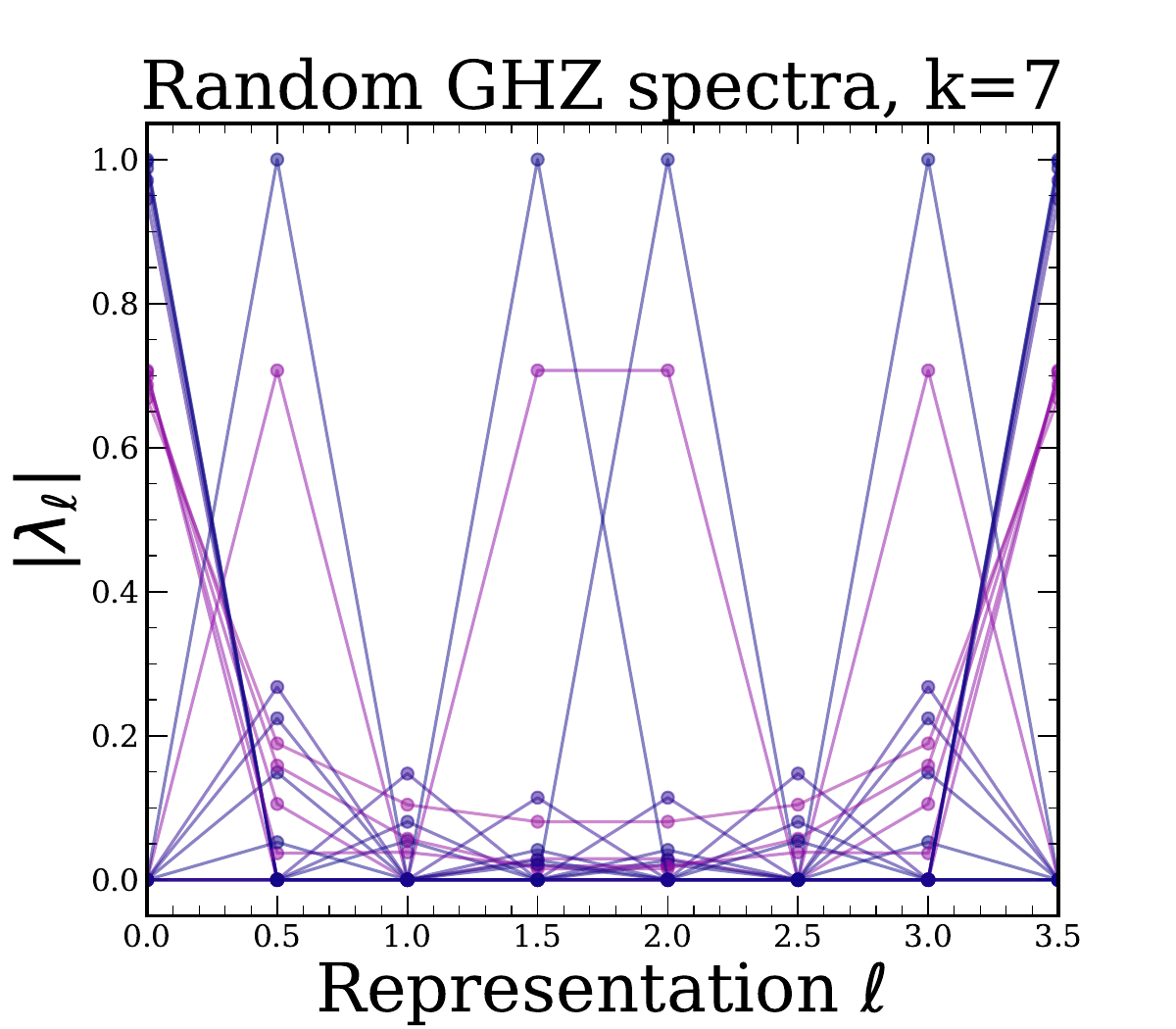}
    \includegraphics[width=0.32\linewidth]{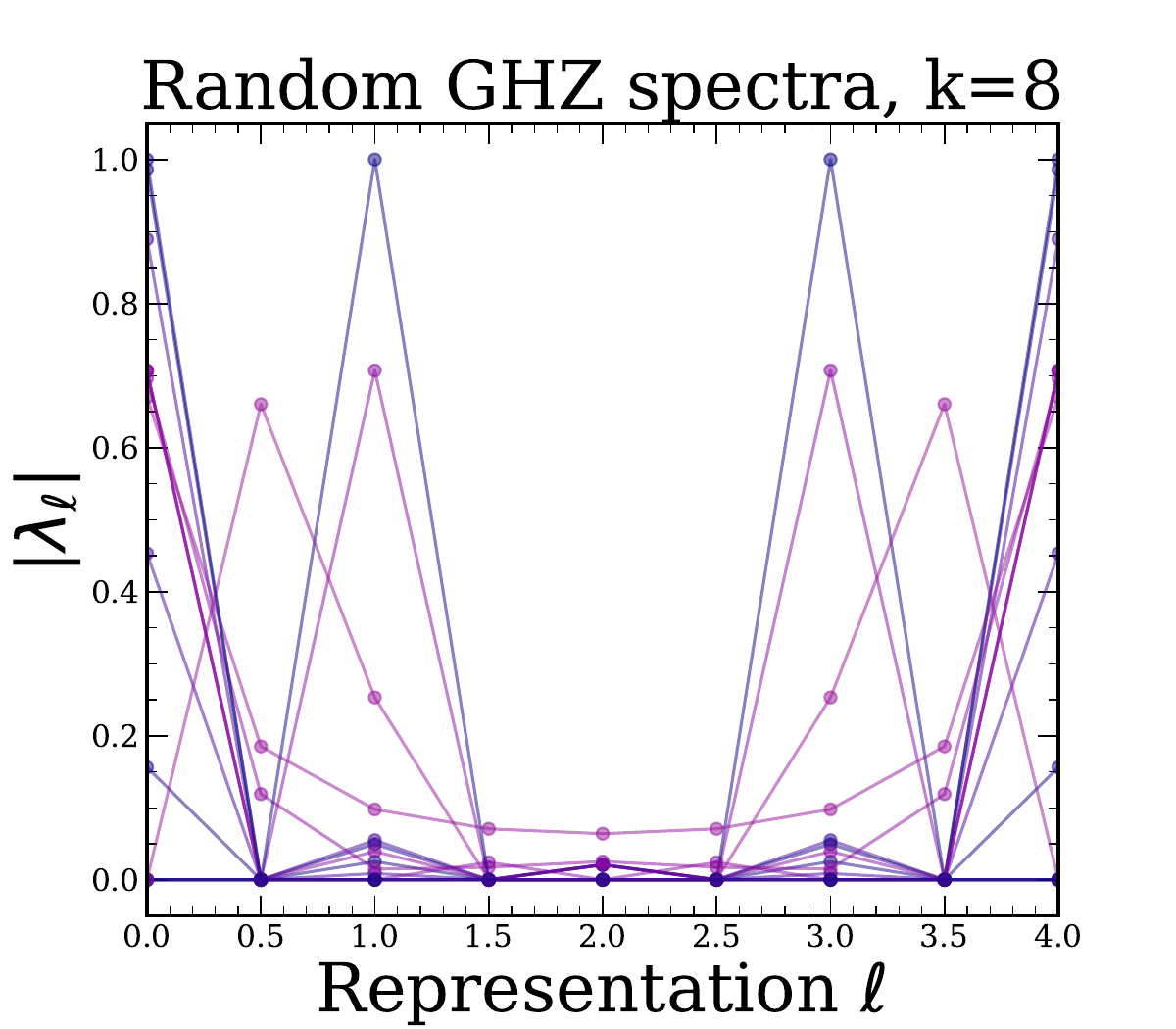}
    \includegraphics[width=0.32\linewidth]{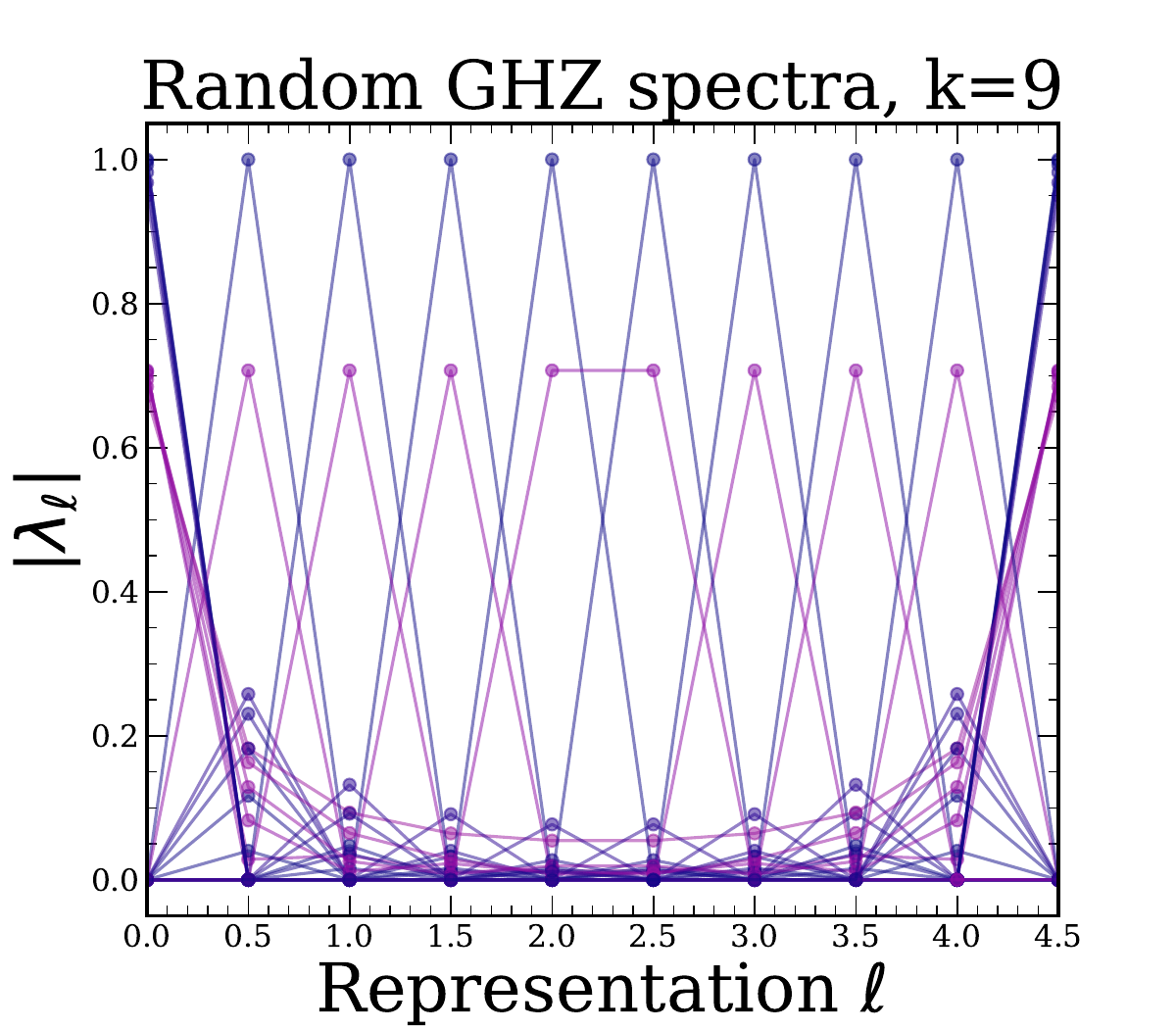}
    \includegraphics[width=0.32\linewidth]{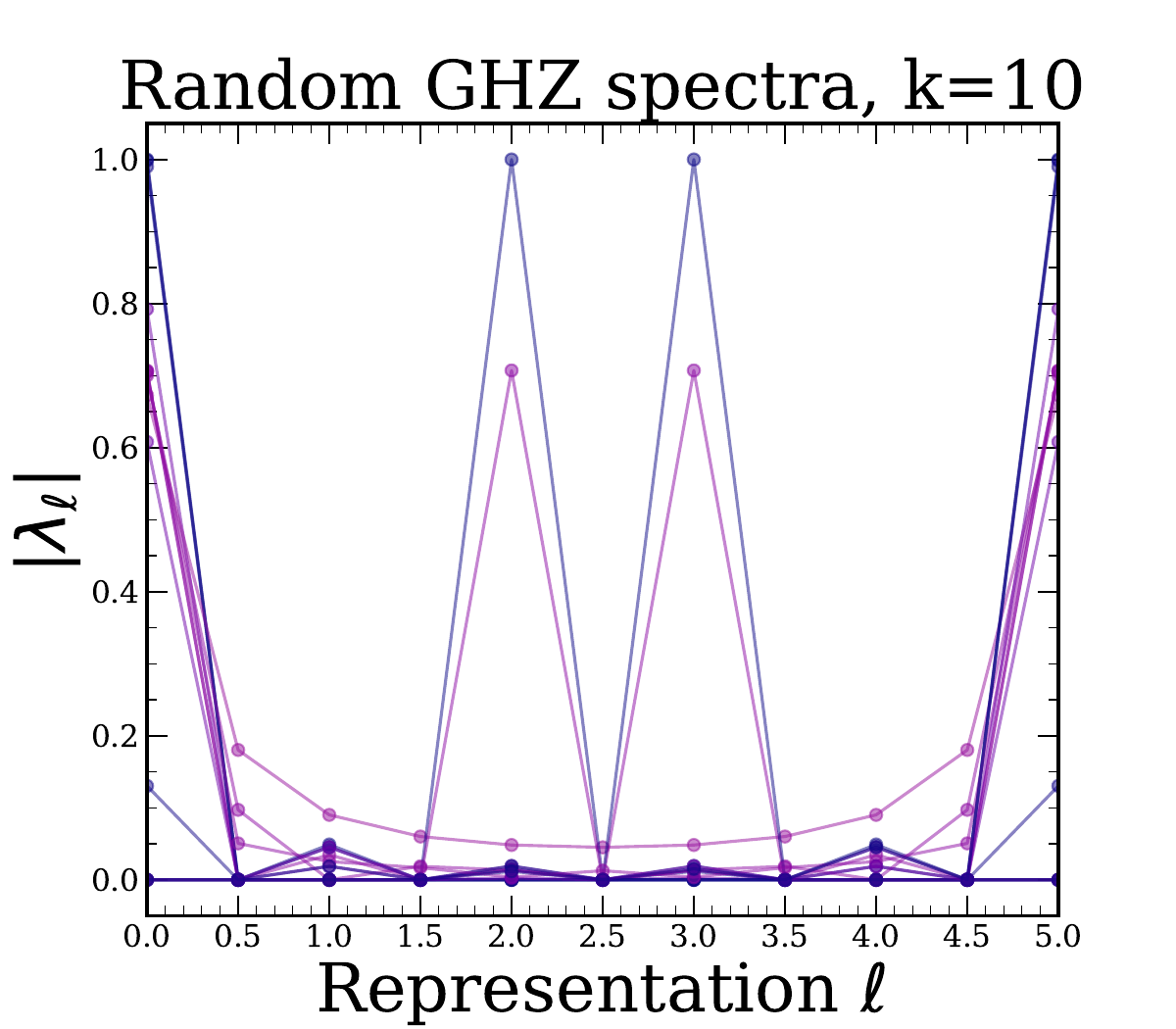}
    \caption{The possible GHZ spectra generated by a single gluing operator $\mathcal{O}$, i.e., for $m=1$, and $\chi=-2$. Lighter colors indicate distributions with a smaller $\max(\lambda_\ell)$, and therefore is an overall flatter distribution than one with a darker color.
    For a given level, higher $m$ GHZ spectra can be generated by multiplying any $m$ of these distributions together after accounting for the overall factor of $(S_{0\ell})^\chi$, and normalizing the result.}
    \label{fig:lambdas}
\end{figure*}

Let $\ket{S}$ be a link state prepared by a Seifert manifold. We saw above that the topology of Seifert manifolds implies that $\ket{S}$ is GHZ-like. In this section, we will investigate to what extent the converse holds: does every GHZ state have a representative of some Seifert manifold? To answer this question, we study the  GHZ spectra $\lambda_\ell$ in more detail. %Our conclusion will be that there is sufficient freedom in the choice of $m$ and $\mathcal{O}^{(i)}$ so that this class of states $\ket{S}$ is dense in the GHZ subspace of $\Ha(T^2)^{\otimes n}$. 

One important observation is that the gluing operators $\mathcal{O}^{(1)}, \cdots, \mathcal{O}^{(m)}$ are in some $k+1$ dimensional representation of $\SL(2,\Z)$. 
This means the matrix elements $\mathcal{O}^{(i)}_{0\ell}$ depend only on the level $k$, the representation $\ell$, and the integers $(\alpha_i,\beta_i)$ in the Seifert symbol. Furthermore, we can make $m$ such choices, one for each operator. Because $m$ is always finite, this means there are only countably many choices of $\mathcal{O}^{(i)}$, and so there are countably many choices of spectrum $\lambda_\ell$. 
This means that, strictly speaking, not every GHZ state in $\Ha(T^2)^{\otimes n}$ arises from the link state of a Seifert manifold, as GHZ states form a continuous (and hence uncountably infinite) family. To that end, there are only a countable number of link states in the boundary Hilbert space $\Ha(T^2)^{\otimes n}$ to begin with.

That being said, how generic should we expect these vectors $\lambda_\ell$ to be? To gain intuition, we investigated the possible spectra $\lambda_\ell$ numerically for $G=\SU(2)$ as follows. Fix the level $k$, $\chi = 2-2g-n$, and $m$. For these parameters, different spectra differ only by the choice of gluing operators $\mathcal{O}$ we use to glue the $m$ solid tori into the manifold. To explore the possible choices of $\mathcal{O}$, we generated $m$ random operators $\mathcal{O}^{(i)}$ by multiplying a random sequences of twists $T$ and $U = TST$, for a total of $N$ twists. As $T$ and $U$ generate $\SL(2,\Z)$, any gluing operator $\mathcal{O}$ can be generated in this way in the limit $N \to \infty$. We took $N$ to be chosen uniformly at random from the interval $[0,1000]$, independently for each operator. This approximately scans the space of possible gluing operators $\mathcal{O}$. Using these operators, we can use \eqref{eqn:GHZspectra} to compute a random GHZ spectrum $\lambda_\ell$. 

We can focus on the $m=1$ case, as these are the GHZ spectra $\lambda_\ell$ which generate the higher $m$ cases.  
Given a sequence of $m=1$ GHZ-spectra $\left\{\lambda_\ell^{(i)}\right\}$, we can use \eqref{eqn:GHZspectra} to generate any higher-$m$ distribution by multiplying them with
\begin{align}
    \lambda_\ell = (S_{0\ell})^{\chi} \prod_{i=1}^m (S_{0\ell})^{-\chi} \lambda_\ell^{(i)} \,.
\end{align}

We generated $10000$ $m=1$ GHZ spectra using the above procedure for $\chi = -2$, and plot the results in Fig.~\ref{fig:lambdas}. These plots have many interesting features. 

\begin{figure}
    \centering
    \includegraphics[width=0.85\linewidth]{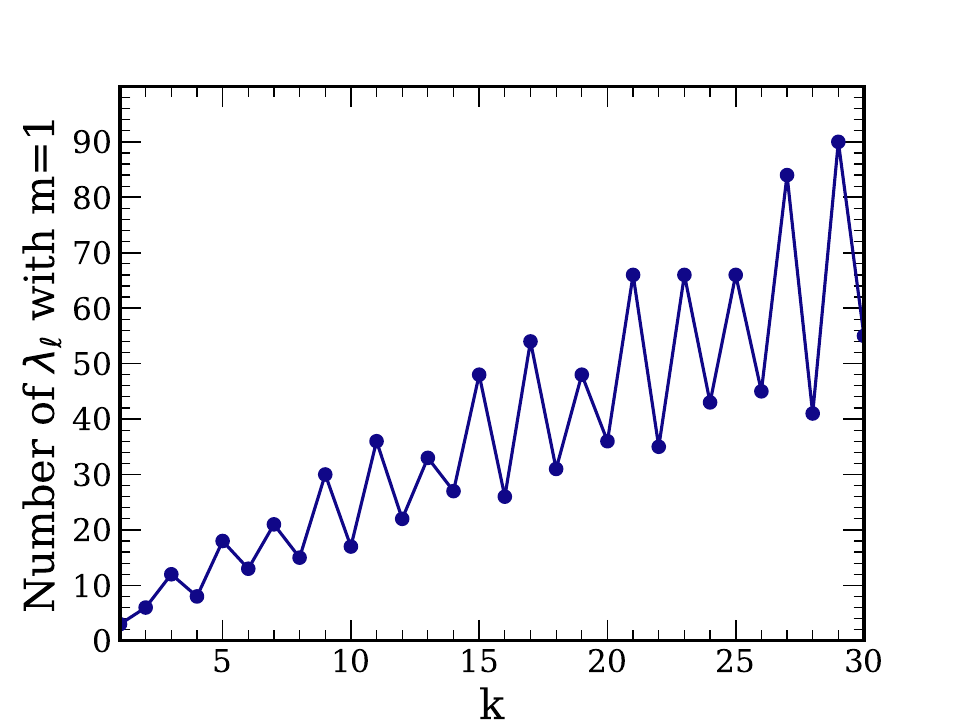}
    \caption{The number of GHZ-spectra $|\lambda_\ell|$ with $m=1$ singular points, as a function of level. Generically, odd levels $k$ have more available distributions than for even levels.}
    \label{fig:countspectra}
\end{figure}

\begin{itemize}
    \item We found that the number of $m=1$ GHZ-spectra is \emph{finite} and level-dependent, which we plot in Fig.~\ref{fig:countspectra}. Indeed, this was proven in \cite{Ng_2010}. Our numerical search is intended to generate this finite list.
    Varying the number of trials from $N=1000$ to $N=10000$, as well as varying the range of samples from $[0,100]$ to $[0,1000]$ to $[0,10000]$ did not change the finite family of distributions we found. This suggests that the finite list of distributions we computed for each level are complete.  
    \item Next, some spectra seem to be ``missing'', such as the delta function on $\ell=1,\frac{5}{2}$ for $k=7$. Furthermore, all even levels are missing delta functions on half integer levels. It would be interesting to understand why certain distributions appear to be missing in more detail.
    \item There is a symmetry $\ell \leftrightarrow \frac{k}{2} - \ell$ acting on the set of possible $m=1$ spectra, even if a particular distribution is not invariant under this flip. This is further evidence that our list of distributions is complete for each level we considered, as there are no distributions with missing partners under this symmetry.
    \item Finally, we can see that for any level, there are GHZ spectra with $\lambda_\ell = \delta_{\ell j}$ for some $j$. Plugging this into \eqref{eqn:periodicGHZstate}, we can see that such states are unentangled. This is a little surprising, as this implies that for a fixed level, up to local unitaries, these Seifert manifolds have an identical Chern-Simons wave function to the unlink. It would be interesting to reverse-engineer the geometry of these manifolds and study them in more detail.
    \item A comment on the above point: a link complement doesn't just produce a Chern-Simons wave function for a fixed level. It produces a choice of state for \emph{any} chosen level. There is no guarantee that a link complement which is unentangled at one level will remain unentangled at a larger level. It would be interesting to better understand the behavior of these ``pseudo-unlinked'' link complements under changing the level of the theory in more detail.
    \item The structure of these $m=1$ distributions suggests that not all GHZ-spectra are achievable using a single periodic link complement. For example, when $k=8$, it is not possible to choose an $m$ such that $\lambda_\ell \sim \delta_{2\ell}$. It would be interesting to classify the subspace of the GHZ subspace of the boundary Hilbert space that periodic link states have access to. 
\end{itemize}

\subsection{Returning to $\mathcal{U}_\phi$}
\label{sec:Uphireturn}

Recall that $\mathcal{U}_\phi$ was the local unitary on the boundary Hilbert space which accounts for the homeomorphism $\phi:\Sigma_{g,n} \to \Sigma_S$ between the model surface $\Sigma_{g,n}$ and the Seifert surface $\Sigma_S$ of the link. Let $U_\phi$ be one of the factors in this local unitary, i.e., $\mathcal{U}_\phi = \otimes_i [U_\phi]_i$. Each $U_\phi$ is an element of $\text{Mod}(1,0) \sim \SL(2,\Z)$, so its matrix elements will be equivalent to one of the $m=1$ gluing operators $\mathcal{O}^{(i)}$ we discussed above. But as shown in Fig.~\ref{fig:countspectra}, for a fixed level $k$, there are a finite number of such operators. This implies that for a fixed level $k$, there are a finite number of possible operators $\mathcal{U}_\phi$, despite there being an infinite number of possible homeomorphisms $\phi$. 

This illustrates another important conceptual difference between $\phi$ and the monodromy $[f]$. Fix the level $k$ of the Chern-Simons theory, and let $\mathcal{C}(\phi)$ denote the minimal number of quantum gates required to construct $\mathcal{U}_\phi$. Because there are a finite number of possible $\mathcal{U}_\phi$, the constant $D = \max_\phi \mathcal{C}(\phi)$ is finite. Thus, link complements $(M_1, \La^n_1)$ and $(M_2,\La^n_2)$ with monodromies conjugate to $[f]$ but differing $\phi$ will differ at most by a constant depth quantum circuit, of depth at most $D$. On the other hand, we can always find link complements whose monodromy $[f],[f']$ are arbitrarily complicated with respect to each other (see \cite{cummings2025entanglementtypicalstateschernsimons} for more information about the complexity of a link's monodromy). This is because, at least for $G = \SU(2)$, $k \geq 9$, $g \geq 2$, and $n=0$, there is a mapping class group element of infinite order \cite{funar2023mappingclassgroupstqft}, so the image of $\text{Mod}(g,0)$ is infinite in these cases. Including the effects of the $n$ punctures adds extra generators to this group, which presumably does not collapse the image from infinite to finite, but we do not know of a direct proof. We expect that adding punctures to increase the complexity of $\text{Mod}(g,n)$ would increase the number of group elements with infinite order in the TQFT representation of the mapping class group. Thus, as far as Chern-Simons theory at fixed level is concerned, there is infinitely more topological data in the monodromy $[f]$ than the homeomorphism $\phi$. \footnote{We thank an anonymous referee for inspiring this discussion.}

%Moreover, the specific group elements which were proven to have infinite image have direct analogs in $\text{Mod}(g,n)$ which no not interact directly with the punctures.

\section{Summary and conclusions}
\label{sec:conclusions}

In this paper, we showed that the monodromy of a fibered link is the link invariant which determines the entanglement structure of the associated link state. We did so by noting the universal factor of $S^1$ in fibered link complements, and canonically quantizing the Chern-Simons path integral along this circle. Canonical quantization led to the appearance of the monodromy operator in \eqref{eqn:monodromyop}. The fact that fibered link states always take the form ``monodromy operator times GHZ state'' proved that the monodromy operator is what allows link states to deviate from GHZ entanglement. We then confirmed that the monodromy controlled the entanglement structure by comparing two infinite families of examples, the Hopf keyring and the Hopf chain. Both families of links can have the same genus $g$ and number of link components $n$. The fact that their entanglement structures are GHZ-like and W-like respectively for $n\geq 4$ was  because their monodromies are different. 

We then used the same formalism for a different slicing of the path integral to prove that all periodic links have GHZ entanglement. This was because all periodic link complements are Seifert manifolds (see Sec.~\ref{sec:seifertmanifold} for a definition of Seifert manifolds, they are not the same as fibering a link complement by their Seifert surfaces). Because Seifert manifolds can be thought of as having ``almost trivial'' monodromy, in the sense that an orbifold base of the manifold has trivial monodromy, they also have GHZ entanglement.

In the  examples we considered, the GHZ structure always had precisely the same origin: the fusion rules of WZW theory. To see this, note that 
\begin{align}
    N_{ijk} = \bra{\chi = -1} \mathcal{S} \ket{ijk}\,,
\end{align}
so even a single fusion matrix had a GHZ state $\ket{\chi=-1}$ of \eqref{eqn:chi_state} built into its definition. Multiplying many fusion matrices together via \eqref{eqn:dimension} computes $\dim(\Ha(g,J))$:
\begin{align}
    \dim(\Ha(g,J)) = \bra{\chi} \mathcal{S} \ket{J}\,.
\end{align}
so the GHZ structure of many fibered link states descends from the fusion rules in a  direct way. For links with local monodromy operators, the relevant GHZ state was the one associated with its Seifert surface. For periodic links, it was the GHZ state associated with the orbifold base of its Seifert manifold (which is not the same as the Seifert surface).

By the Nielsen-Thurston classification, there are only two remaining types of links: hyperbolic and reducible. The obvious next step is to analyze hyperbolic and reducible links in more detail. Based on the examples $K_n,H_n$, we can see that the possibilities for multipartite entanglement will generally be richer than pure GHZ entanglement in these cases. 

Finally, many of these results can be immediately generalized to link states in many other three dimensional TFTs. 
The only feature of Chern-Simons theory we used in this paper was that it has a chiral rational CFT (RCFT) dual; namely, the conformal blocks of the associated WZW theory.
Using the Turaev-Viro \cite{TVTFT} (partition function) or Reshetikhin-Turaev \cite{Reshetikhin:1991tc} (wave functions) constructions, we can begin with any chiral RCFT and build an associated three dimensional TFT. 
Examples of TFTs which can be built using these methods include Dijkgraaf-Witten theories \cite{Dijkgraaf:1989pz,petit2006invariantturaevvirogroupcategory}, Levin-Wen models \cite{Levin_2005,Kirillov:2011mk}, and Chern-Simons theory itself \cite{Andersen:2012gs}.
Our results can be immediately applied to link states of these TFTs, and the monodromy $[f]$ still controls their entanglement structure in these cases.\footnote{We thank Jonathan Heckman for pointing this out.} 

\paragraph{Acknowledgments:} We are grateful to Qingyue Wu, Benjamin Bode, Chitraang Murdia, Onkar Parrikar, and Rob Leigh for useful discussions. Additionally, we thank Shaunak Modak and Nishant Mishra for helpful comments about the numerics of random periodic link states. It is also a great pleasure to thank Paul Severino, who introduced CC to the machinery of Seifert surfaces in the course of many helpful discussions. CC is supported by the National Science Foundation Graduate Research Fellowship under Grant No. DGE-2236662.   VB is supported in part by the DOE through DE-SC0013528 and QuantISED grant DE-SC0020360, and in part by the Eastman Professorship at Balliol College, University of Oxford.

\appendix

\section{Integrable representations of a group $G$ at level $k$}
\label{sec:intreps}

Our discussion in this appendix essentially follows chapter 14 of \cite{DiFrancesco:1997nk}.

For the moment, let us discuss the theory of representations of Lie groups, which formally corresponds to the classical limit $k \to \infty$.  An important fact about Lie groups is that their representation theory is well understood, providing concrete models where their properties can be investigated. To this end, the most useful kind of representation of a Lie group are those which are finite dimensional. These representations are most systematically understood by first constructing representations of the Lie algebra $\mathfrak{g}$ of $G$ and exponentiating them to a full $G$ representation. For instance, consider $\mathfrak{g} = \mathfrak{su}(2)$. One can build any finite dimensional representation of $\mathfrak{su}(2)$ by first declaring a ``highest weight'' state $\ket{j,+j}$ which is annihilated by the raising operator $J^+$.
Any other state in this representation can be built by acting appropriate powers of $J^-$ on this highest weight state, so a choice of $j$ uniquely determines the representation. 
This procedure leads to the complete spin-$j$ representation, where $j$ is any integer or half integer. 

A finite dimensional representation of a more complicated Lie algebra proceeds in a similar way. Instead of picking a single (half) integer $j$, we must pick a $r$-dimensional vector $\vec{\lambda}$, called the \emph{Dynkin labels} for the representation. Here, $r$ is the \emph{rank} of the group. The rank of a Lie group is equal to the number of nodes on its Dynkin diagram, and for $\su(N)$, $r = N-1$. Thus, for $\mathfrak{su}(2)$, $r=1$, explaining why their representations are indexed by a single number, the spin quantum number.  From this Dynkin label, we can associate it with a ``highest weight'' representation, where $\vec{\lambda}$ plays a similar role to $j$ in the $\su(2)$ case. Instead of a single creation operator $J^+$, there are now $r$ different creation operators $E^\alpha$ which annihilate $\big|{\vec{\lambda}}\big\rangle$. It is common to leave the Dynkin label implicit, so $\big|{\vec{\lambda}}\big\rangle$ refers to what we might have called $\big|{\vec{\lambda}}\big\rangle$ in analogy to $\ket{j,+j}$. Similar to how $J^- \ket{m} \sim \ket{m-1}$ for $m > -j$, each anihilation operator $(E^\alpha)^\dagger$ acts on the highest weight state via $(E^\alpha)^\dagger \big|{\vec{\lambda}}\big\rangle \sim \big|{\vec{\lambda}}\big\rangle$, with the condition that $\big|{\vec{\lambda}}\big\rangle = 0$ if $\vec{\lambda} - \vec{\alpha}$ is not in the associated lattice of states in the representation. Each vector $\vec{\alpha}$ associated with a creation operator $E^\alpha$ is called a root of $\mathfrak{g}$. The $E^\alpha$ are various linear combinations of the Lie algebra generators, and their commutation relations follow directly from that of the Lie algebra. Every state in the highest weight representation associated to $\vec{\lambda}$ can be built from acting the annihilation operators $(E^\alpha)^\dagger$ on $\big|{\vec{\lambda}}\big\rangle$. 

Because we took $\vec{\lambda}$ to be a highest weight representation, its components $\lambda_i$, $i = 1, \cdots r$ must all be positive. These components are called Dynkin labels. It turns out that by switching on a finite level, the effect is to introduce a new Dynkin label $\lambda_0 = k - \sum_{i=1}^r \lambda_i a^\vee_i $, where $a^\vee_i$ are integers called the comarks of their associated root $\vec{\alpha}_i$. For the case of $\su(N)$ (or any other simply laced Lie algebra), $a^\vee_i = 1$ for all $r$ roots.
The constraint that $\lambda_i \geq 0$ for $i = 0, \cdots, r$ means that there is a finite, level dependent amount of representations which are allowed. Note that as $k \to \infty$, any positive $\lambda_i$ are allowed, recovering the case of a classical Lie algebra.

\section{When is a link fibered?}
\label{sec:whenfibered}

In this appendix, we review some of the methods to recognize if a link is fibered.

\subsection{The Alexander polynomial}

This brief review of the Alexander polynomial will be developed just enough to explain why it is sometimes a useful invariant for telling when a link is fibered. For a more complete review, we refer the reader e.g. to \cite{rolfsen1976knots}.

Link invariants are useful because they can be used to distinguish between different links. Often in knot theory, this comes in the form of various recursively-defined polynomials. One example is the colored Jones polynomial, famously computable using the Wilson loop correlation functions of Chern-Simons theory. A different example of a polynomial invariant, which we will now focus on in this section, is the \emph{Alexander polynomial}. One definition of the Alexander polynomial $\Delta_\La(t)$ of a link $\La$ is the recursive skein relation
\begin{align}
    \Delta_{\text{unknot}}(t) &= 1 \,,\\
    \Delta_{\La_+}(t) + (t^{1/2} - t^{-1/2}) \Delta_{\La_0}(t) - \Delta_{\La_-}(t) &= 0\,.
\end{align}
Here, $\La_+,\La_0,\La_-$ are links that differ from $\La$ by only a single crossing, the difference being determined by the subscript. The Alexander polynomial is known for many knots and links for sufficiently low crossing. The Alexander polynomial is more ``coarse'' than the Jones polynomial, because e.g. it can not distinguish a link from its mirror image. Despite this, the Alexander polynomial is a commonly used link invariant in knot theory for two reasons. For one, it is easy to compute. Secondly, and more importantly for us, it has a nice interpretation if the link is fibered. In that case, there is another link invariant which we have seen many times in the main text: the link monodromy. The Alexander polynomial is deeply related to the link monodromy in the following way.

Let $\La$ be a fibered link with a Seifert surface $\Sigma'_S$ which is \emph{not} necessarily of minimal genus. As $\Sigma'_S$ is a two dimensional surface, we can define its integral homology group $H_1(\Sigma'_S,\Z)$ in a straightforward way. With the cobordism picture of link complements in mind, which we described in Sec.~\ref{sec:math}, we would like to understand what happens to a basis of $H_1(\Sigma_S,\Z)$ after this cobordism takes place. If we think of splitting the link complement along $\Sigma'_S$ so that the (probably non-trivial) cobordism is between two copies of $\Sigma'_S$, we can analyze this by comparing generators of $H_1(\Sigma'_S,\Z)$ which are just ``below'' and ``above'' $\Sigma'_S$. This would correspond to curves at the beginning and end of this cobordism, and hence capture the effect of passing a generator of $H_1(\Sigma'_S,\Z)$ along. Since $H_1(\Sigma'_S,\Z)$ is a vector space, we would like to understand this transformation as a linear map $V:H_1(\Sigma'_S,\Z) \to H_1(\Sigma'_S,\Z)$, which we call the \emph{Seifert matrix}. The Seifert has matrix elements
\begin{equation}
   V_{ij} = \text{lk}(e_i,e_j^+) \,.
\end{equation}
In words, $V_{ij}$ is the integer which computes the linking number between a basis element $e_i$ and another basis element $e_j$ after it has been pushed around the $S^1$ of the link complement. The $+$ superscript is just indicating that we take $e_j$ to be on one side of a thickened neighborhood of $\Sigma'_S$, where we take $+$ to be defined by the orientation of $\Sigma'_S$\footnote{We are greatly simplifying this story, and ignoring some details about universal covering spaces of link complements that would take us too far afield. We refer to interested reader to the Rolfsen for more complete details.}.

The Seifert matrix is \emph{not} a link invariant, as it can change when we change our (arbitrary) choice of basis for $H_1(\Sigma_S,\Z)$. However, the polynomial
\begin{align}
    \Delta_\La(t) = \det(V - t V^T)
\end{align}
\emph{is} a link invariant, up to multiplication by $\pm t^{\pm m}$ for some $m \in \Z$. This ambiguity arises because we did not demand that $\Sigma'_S$ was minimal genus.
This ambiguity is the \emph{only} ambiguity of $\Delta_L(t)$, and it can be fixed by demanding that the lowest order term is a positive constant. This is an alternative definition for the Alexander polynomial. 

If the link is fibered, then $\Delta_\La(t)$ has extra constraints. First, we note that because $V$ captures the effect of sending curves around the link complement cobordism, it is intimately related to the link monodromy. In particular, if $\La$ fibers with Seifert surface $\Sigma_S$ and monodromy $[f]$, and $f_*:H_1(\Sigma_S,\Z) \to H_1(\Sigma_S,\Z)$ is the induced map on homology by the link monodromy (which is invertible because $[f]$ is), then
\begin{align}
    \Delta_\La(t) = \det(t - f_*) \,.
\end{align}
Thus, a basic requirement for fibered links is that the leading coefficient of $\Delta_\La(t)$ is $\pm 1$. If $\La$ is an alternating knot, then this is also a sufficient condition, but generally it is not sufficient. Because of this, the Alexander polynomial is more useful as determining a possible obstruction to a link being fibered, rather than determining when the link is fibered. At the same time, \emph{any} polynomial is the Alexander polynomial of some link \cite{stallings}. So there are infinitely many links which are not fibered, because their Alexander polynomial does not have a leading coefficient of $\pm 1$. 

\subsection{Homogeneous braids}

A useful way to construct links in $S^3$ is via their presentation as a \emph{braid closure}. We will not go into all the details about the braid group here, but refer the reader to \cite{rolfsen1976knots} for more details about the application of the braid group to knot theory.

The class of braids which we wish to discuss in this section are \emph{homogeneous braids}. A homogeneous braid is a braid which contains only positive (or only negative) crossings. A link whose braid representation is homogeneous is called a homogeneous link. Examples of homogeneous links include torus links, but non-examples include the Borromean rings. In \cite{stallings}, Stallings proved that all homogeneous links are fibered links. The converse is not necessarily true, because e.g. the Borromean rings are fibered but not homogeneous.
Because homogeneous links are fibered, they have a well defined monodromy, and the monodromy of homogeneous links is discussed extensively in \cite{bell2012monodromieshomogeneouslinks}.

We can use the fact that all homogeneous links are fibered to prove the claim in Sec.~\ref{sec:nonfibered} that any non-fibered link $\La$ in $S^3$, has a fibered superlink $\La'$, also in $S^3$. Thinking of $\La$ as being the closure of a braid $B$, Stallings showed this through the following steps.

\begin{itemize}
    \item An arbitrary braid $B$ can be made homogeneous by introducing additional (non-unique) strands to form a new braid $B \cup S$. We think of $\La'$ as being the closure of $B \cup S$. Generically, the braid closures of $B$ and $B \cup S$ are not the same, so $\La$ and $\La'$ are not the same link. By Stalling's result, $\La'$ is a homogeneous, fibered link. 
    \item We can arrange these additional strands $S$ so their closure is disjoint from the original braid.
    This is what ensures that the superlink $\La'$ contains $\La$ as a sublink.
    \item We can introduce additional positive crossings on the $S$ strands so that they close to a single extra component. This is what ensures that $\La' = \La \cup K$ for some knot $K$, linked to $\La$. So far, the linking numbers of $K$ with $\La$ and itself are still arbitrary.
    \item We can choose the self-linking number of $K$ to vanish. This is what ensures that the additional component $K$ is an unknot.
    \item Finally, we can pick the linking numbers of $K$ and $\La$ so that they sum to $1$. This is what guarantees that $\La'$ fibers in $S^3$ \cite{stallings}.
\end{itemize}

Thus, we have shown that any link $\La$ in $S^3$ is a sublink of some fibered link $\La' = \La \sqcup K$, where $K$ is an unknot. This implies that the methods we develop in this paper can be applied to arbitrary links, with the additional steps that we explain in Sec.~\ref{sec:nonfibered}.

\bibliographystyle{hunsrt}
\bibliography{biblio}

\end{document}